\documentstyle[eqsecnum,aps,multicol]{revtex}
\begin{document}
\def\s.{\sigma}
\def\be.{\begin{equation}}
\def\ee.{\end{equation}}
\def\la.{\langle}
\def\ra.{\rangle}
\def\st.{\tilde{\Sigma}}
\def\tw.{\tilde{\omega}}
\def\ha.{\hbox{\tiny A}}
\def\hb.{\hbox{\tiny B}}
\def\haa.{\hbox{\tiny AA}}
\def\hbb.{\hbox{\tiny BB}}
\def\da.{\downarrow}
\def\ua.{\uparrow}
\def\w.{\omega}
\def\di.{d^\infty}
\def\cg.{{\cal G}}

\title{Insulating Phases of the $d=\infty$ Hubbard model}
\author{David E Logan, Michael P 
Eastwood and Michael A Tusch}
\address{Physical and Theoretical Chemistry Laboratory, 
University of Oxford, South Parks Road, Oxford OX1 3QZ (U.K.) }

\maketitle

\begin{abstract}
 A theory is developed for the $T=0$ Mott-Hubbard insulating phases 
of the $d^\infty$ Hubbard model at $\frac{1}{2}$-filling, including both the
antiferromagnetic (AF) and paramagnetic (P) insulators. Local
moments are introduced explicitly from the outset, enabling
ready identification of the dominant low energy scales for
insulating spin-flip excitations. Dynamical coupling of 
single-particle processes to the spin-flip excitations leads to a
renormalized self-consistent description of the single-particle
propagators that is shown to be asymptotically exact in strong
coupling, for both the AF and P phases. For the AF case, the
resultant theory is applicable over the entire $U$-range, and is
discussed in some detail. For the P phase, we consider in
particular the destruction of the Mott insulator, the resultant
critical behaviour of which is found to stem inherently from
proper inclusion of the spin-flip excitations.

\end{abstract}


\begin{multicols}{2}
\narrowtext

\section*{1. Introduction}    
\addtocounter{section}{1}                 

  Since its inception more than thirty years ago \cite{r1}, the
Hubbard model has become the canonical model of interacting
fermions on a lattice. Although possibly the simplest model
to describe competition between electron itinerancy and
localization, with attendant implications for a host of
physical phenomena from magnetism to metal-insulator
transitions, its simplicity is superficial and an exact
solution exists only for $d=1$ dimension \cite{r2}.

  Recently, Metzner and Vollhardt \cite{r3} have pointed to the
importance of the opposite extreme, $d=\infty$. In suppressing
spatial fluctuations, the many-body problem here simplifies
considerably, reducing to a dynamical single-site mean-field
problem. Motivated in part by the expectation that an
understanding of the $d=\infty$ limit will serve as a starting point
for systematic investigation of finite dimensions, and by the
knowledge that some important vestiges of finite-$d$ behaviour
remain inherent in the $d=\infty$ limit, intense study of the
$\frac{1}{2}$-filled $d=\infty$ Hubbard model on bipartite lattices has since
ensued; for recent detailed reviews, see Refs \cite{r4,r5,r6}.  
 
  The true ground state of the model is an antiferromagnet (AF)
for all interaction strengths $U>0$. One aim of the present work
\cite{r7} is to develop a theory for the $d=\infty$  AF which, in contrast 
to previous theories for the AF phase \cite{r8,r9,r10}, is reliable over the
entire $U$-range, and in particular becomes exact in the
$U\rightarrow\infty$  strong
coupling limit both at $\frac{1}{2}$-filling where the Hubbard model maps onto
the AF Heisenberg model, and in the one-hole sector where it reduces 
to the $t$-$J$ model \cite{r11}

  The majority of previous work \cite{r4,r5,r6} on the $d=\infty$  Hubbard model has
focused on the paramagnetic (P) phase that results, even for $T=0$,
simply by neglecting the magnetic ordering (or suppressing it via
frustration \cite{r5}). One highlight of this work has been the emergence
of a detailed description of the Mott metal-insulator transition,
although here too the picture is not complete: for example, a firm
understanding of the mechanism by which the $T=0$ Mott insulating
solution is destroyed, and even whether it is continuous or first-order,
remains elusive \cite{r5,r12}. A second aim of this paper is to focus on the 
insulating state of the P phase, and to develop a theory for it on a
footing essentially identical to that for the AF, which likewise becomes
exact in strong coupling and which permits an analysis of the destruction
of the Mott insulator.

  In seeking to develop a `unified' description of the AF and P insulating
phases, we adopt a rather different approach to that taken in previous
work \cite{r4,r5,r6} by introducing explicitly, and from the outset, the notion of
site local moments. To this end we consider first a conventional $T=0$
mean-field approach to the problem in the form of unrestricted 
Hartree-Fock (UHF), together with a random phase approximation (RPA) 
for transverse
spin excitations of the mean-field state. Despite the limitations of such
an approach {\em per se}, its importance resides in enabling identification
of key low energy scales for insulating spin-flip excitations. Since
spatial fluctuations are suppressed for $d=\infty$, the low energy spin-flip
excitations are found to be Ising-like and (for each phase) 
characterized by a single scale, $\omega_s$. This has a simple physical
interpretation. For the AF, $\omega_s=\omega_p(U)$  is essentially just the energy
cost of flipping a spin in the N\'eel ordered background; the ubiquity
of antiferromagnetism for all $U>0$ leads naturally to $\omega_p>0$ for all $U$,
with $\omega_p\sim 1/U$ as $U\rightarrow\infty$ as one expects in the
Heisenberg limit. For the
P phase by contrast, where magnetic ordering is absent, the fact that
a given spin is equally likely to be surrounded by $\ua.$- or $\da.$-spins and
thus (for $d=\infty$) has as many $\ua.$- as $\da.$-spin neighbours, 
ensures that the 
corresponding spin-flip energy cost $\omega_s=0$ for all $U$ in the insulating
state.

  Identification of the low energy spin-flip scales, while crucial to
the present work, is preliminary: to transcend the limitations of
the conventional `static' mean-field approach, single-particle processes
must subsequently be coupled dynamically to the transverse spin-flip
excitations. It is this which, in leading as we shall describe to a
self-consistent description of the single-particle Green functions,
enables the aims outlined in the preceding paragraphs to be achieved.

  The Hubbard Hamiltonian, in standard notation, is
  
\be.
H=-t\sum_{(ij)\s.}c_{i\s.}^\dagger c_{j\s.}+U\sum_i
n_{i\ua.}n_{i\da.}~~~~~: t=t_\ast/\sqrt{2Z}
\ee.
with the $(ij)$ sum over nearest neighbour sites on a bipartite lattice 
of coordination number $Z$: a Bethe lattice (on which in practice we
shall largely focus), or a d-dimensional hypercube. To ensure a
non-trivial limit as $d\rightarrow\infty$ \cite{r3}, the hopping is
scaled as $t = t_\ast/\sqrt{2Z}$.
The paper is organised as follows. UHF+RPA, and the spin-flip scales 
referred to above, are discussed in \S 2. Emphasis is also given here to
simple physical arguments which, in highlighting both the deficiencies 
and virtues of UHF+RPA, indicate what is required to go beyond it;
particular attention being given in this regard to UHF for the P phase,
in view of its close relation to the early work of Hubbard \cite{r13} and the
Falicov-Kimball model \cite{r14}.

  Dynamical coupling of single-particle processes to the transverse
spin excitations is considered in \S 3, leading (\S 3.1) to a renormalized
self-consistent approximation for the ($T=0$) single-particle Green functions
upon which we subsequently concentrate. In \S 3.2 the strong coupling
behaviour is examined analytically, and shown to be asymptotically 
exact for both the P and AF phases. Results are given in \S 3.3, focusing
in particular on single-particle spectra for the AF from strong to weak
coupling, and on a discussion of the localization characteristics of
the single-particle excitations --the latter being quite subtle, and
pointing to the delicacy of the limit $U\rightarrow\infty$ for the AF phase. For the
P insulator, single-particle spectra are discussed briefly in \S 3.3,
before considering the destruction of the Mott insulating solution in
\S 4. The single-particle gap is found to close continuously, with an
exponent $\nu=1$, at a critical $U_c=3.41t_\ast$. The origins of this
behaviour are found to stem from inclusion of the $\omega_s=0$
spin-flip scale in the interaction
self-energies, pointing 
to the importance of such throughout the entire insulating regime, and
not solely in obtaining the exact strong coupling limit. The results of
\S 4 are in good agreement with recent numerical work \cite{r12}, as discussed
in \S 5.

\section*{2. Conventional mean-field approach}
\addtocounter{section}{1}
\setcounter{equation}{0}

We focus on the zero temperature single-particle Green functions,
defined by
\be.
G_{ii;\s.}=-i\la.T\{c_{i\s.}(t)c_{i\s.}^\dagger\}\ra.\equiv G_{ii;\s.}^+(t)+G_{ii;\s.}^-(t)
\label{21}
\ee.
(for the site diagonal element); and separated for later purposes into
retarded ($+,t>0$) and advanced ($-,t\leq 0$) components. The essential
feature of $d^\infty$ is that the corresponding interaction
self-energy is site-diagonal \cite{r3,r15},
$\st._{ij;\s.}(\tw.)=\delta_{ij}\st._{i\s.}(\tw.)$; here, and
throughout, $\tw.$ denotes frequency relative to the Fermi level, viz
$\tw.=\omega-U/2$. $G_{ii;\s.}(\tw.)$ may be written as
\begin{mathletters}
\label{22}
\be.
G_{ii;\s.}(\tw.)=[\tw.-\st._{i\s.}(\tw.)-S_{i\s.}(\tw.)]^{-1}
\ee.
where $S_{i\s.}$ is the `medium' self-energy ---which alone survives
in the non-interacting limit--- expressing hopping of $\s.$-spin electrons
to neighbouring sites. Simple application of Feenberg's renormalized
perturbation theory \cite{r16,r17} shows that, for $d=\infty$ but regardless
of lattice type, $S_{i\s.}$ is a functional solely of the
$\{G_{jj;\s.}\}$. The functional dependence is particularly simple for
the Bethe lattice (BL) on which we concentrate, namely
\be.
S_{i\s.}(\tw.)=\sum_jt_{ij}^2G_{jj;\s.}(\tw.)
\ee.
\end{mathletters}
with $t_{ij}=t_\ast/\sqrt{2Z}$ the nearest neighbour hopping
element. Note that this is quite general; no assumption has been made
about magnetic ordering or otherwise.

We consider now a conventional mean-field approach to the
single-particle Green functions.

\subsection*{2.1 UHF}

For both the AF and P phases, a Hartree-Fock approximation ---by which
we emphasize is here meant spin {\em unrestricted} Hartree-Fock
(UHF)--- is the simplest non-trivial mean field approximation, in
which the notion of site local moments ($\mu_i$), regarded as the
first effect of electron interactions, enters from the outset. In the
AF case, the local moments are naturally ordered in an A/B
2-sublattice N\'eel state, with $\mu_i=\pm |\mu|$ for site $i$ in the
A/B sublattice respectively \cite{r18}. For the P phase by
contrast, the local moments are randomly oriented: a site is equally
likely to be A-type as B-type \cite{r19}. In either case the
essential ---and limiting--- feature of UHF is that it is a static
approximation, with solely elastic scattering of electrons and
$\omega$-independent interaction self-energies approximated by
\be.
\label{23}
\st._{\hbox{\tiny A}\s.}^0=-\frac{\s.}{2}U|\mu|=-\st._{\hbox{\tiny B}\s.}^0~~.
\ee.

For the AF phase, the UHF Green functions ($G_{ii;\s.}^0\equiv
G_{\alpha\s.}^0$ with $\alpha=$A or B) follow from Eqs (\ref{22},\ref{23}) for the
BL as 
\be.
\begin{array}{ll}
G^0_{\ha.\s.}(\tw.) &=[\tw.+\frac{\s.}{2}U|\mu| 
-\frac{1}{2}t_\ast^2G_{\hb.\s.}^0(\tw.)]^{-1} \cr
G^0_{\hb.\s.}(\tw.) &=[\tw.-\frac{\s.}{2}U|\mu| 
-\frac{1}{2}t_\ast^2G_{\ha.\s.}^0(\tw.)]^{-1} 
\end{array} \hspace{1.0cm} \hbox{: ~AF}
\label{24}
\ee.
where the `medium' self-energy part reflects the 2-sublattice
structure of the N\'eel state. Eqs (\ref{24}) are a closed set, with
the UHF local moment $|\mu|=|\mu_0|$ found self-consistently via the
usual gap equation (see e.g. \onlinecite{r20}), which may be written
formally as 
\be.
|\mu_0|=\int_{-\infty}^0d\tw. 
\ \left(D_{\ha.\uparrow}(\tw.)-D_{\ha.\downarrow}(\tw.)\right)
\label{25}
\ee.
in terms of the corresponding spectral densities. And the total Green
function is given by
\be.
G^0(\tw.)={\scriptstyle\frac{1}{2}}[G^0_{\ha.\s.}(\tw.)+G^0_{\hb.\s.}(\tw.)]
\label{26}
\ee.
such that $D^0(\tw.)=-\pi^{-1}{\rm sgn}(\tw.){\rm Im}G^0(\tw.)$ gives
the total single-particle spectrum.

For the P phase  by contrast,
\be.
\begin{array}{ll}
G^0_{\ha.\s.}(\tw.)& =[\tw.+\frac{\s.}{2}U|\mu| 
-\frac{1}{2}t_\ast^2G^0(\tw.)]^{-1} \cr
G^0_{\hb.\s.}(\tw.)& =[\tw.-\frac{\s.}{2}U|\mu| 
-\frac{1}{2}t_\ast^2G^0(\tw.)]^{-1} 
\end{array}\hspace{1.0cm}\hbox{: ~P}
\label{27}
\ee.
The sole difference to Eq. (\ref{24}) occurs in the medium self-energy
(see Eq. (\ref{22}b)), since the nearest neighbours to any site are
equally likely to be A- or B-type sites. Eqs (\ref{26}) are a closed
set for $G^0(\tw.)$ and the $G^0_{\alpha\s.}(\tw.)$ in the P phase; the UHF
local moment is again found from Eq. (\ref{25}).

For either phase there are two basic symmetries, viz
\begin{mathletters}
\label{28}
\begin{eqnarray}
D_{\ha.\s.}^0(\tw.)&=&D_{\hb.-\s.}^0(\tw.)\\ 
&=& D_{\ha.-\s.}^0(-\tw.)
\end{eqnarray}
\end{mathletters}
reflecting the $\uparrow/\downarrow$-spin symmetry
($G^0_{\ha.\s.}(\tw.)=G^0_{\hb.-\s.}(\tw.)$) and particle-hole symmetry
($G^0_{\ha.\s.}(\tw.)=-G^0_{\ha.-\s.}(-\tw.)$) respectively; and note
therefore from Eq. (\ref{26}) that $G^0(\tw.)$ is naturally
independent of the spin, $\s.$.

We add further that UHF yields the correct atomic limit (where
$|\mu_0|=1$) for either phase, as is clear from Eqs
(\ref{24},\ref{26},\ref{27}) with $t_\ast=0$.

\subsubsection*{A. Antiferromagnet}

UHF for the AF has been widely studied since the early work of Penn
\cite{r18}. Here we mention only that for any $d>1$ the exact ground
state of the $\frac{1}{2}$-filled Hubbard model on a bipartite lattice
is an AF insulator for all $U>0$, and this is qualitatively well
captured at UHF level: for all $U>0$, the mean-field ground state is a
2-sublattice N\'eel AF, with a gap in the single-particle spectrum
$D^0(\tw.)$ given by $\Delta(U)=U|\mu|$; Fig. 1 shows $D^0(\tw.)$ at
$U/t_\ast=4$ for the $d^\infty$ BL.

The deficiencies of UHF are however most clearly seen in strong
coupling, $U\rightarrow\infty$, where for the AF the single-particle
spectrum reduces to
$D^0(\tw.)=\frac{1}{2}[\delta(\tw.+\frac{U}{2})+\delta(\tw.-\frac{U}{2})]$
---as for the atomic limit, $t_\ast=0$. The physical origin of this is
simple: consider for example the upper Hubbard band in strong
coupling, and imagine adding an $\uparrow$-spin electron to a site
(B-type) already occupied by a $\downarrow$-spin. Since UHF is an {\em
independent} (albeit interacting) electron approximation, only the
added $\uparrow$-spin can potentially hop to nearest neighbour (NN)
sites. But it cannot do so in the strong coupling limit, since for the
AF all NN's to the $\downarrow$-spin B-site are $\ua.$-spins
(A-type). The added $\ua.$-spin thus effectively `sees' the
$\da.$-spin site as an isolated site, hence the emergence of atomic
limit behaviour as the strong coupling limit at UHF level. But while
physically transparent, this behaviour is wrong. In strong
coupling, and for the 1-hole sector appropriate to the lower Hubbard
band (or 1-doublon sector for the upper Hubbard band), the Hubbard
model maps onto the $t$-$J$ model\cite{r11}
\be.
\hat{H}_{tJ}=-t\sum_{(i,j)\s.}\bar{c}_{i\s.}^\dagger\bar{c}_{j\s.}
+{\scriptstyle\frac{1}{2}}J_\infty\sum_{(
i,j)}\left(\bbox{S}_i\cdot \bbox{S}_j-{\scriptstyle\frac{1}{4}}
n_i  n_j\right)
\label{29}
\ee.
where  the hole moves in a restricted subspace of no doubly occupied
sites ($\bar{c}_{i\s.}^{\dagger}=c_{i\s.}^\dagger(1-n_{i-\s.})$); and in
the fluctuating spin background provided by the Heisenberg part of
$\hat{H}_{tJ}$, with NN exchange coupling $J_\infty=4t^2/U$. Although
it is exact in the atomic limit, UHF by itself can evidently say essentially
nothing about the strong coupling limit.

\subsubsection*{B. Paramagnet}

UHF for the $T=0$ paramagnetic phase warrants separate discussion, in
part because of its very close relation to two other well known
approaches. The first is that due to Hubbard \cite{r13}, with
`spin-disorder scattering' only. This is often called the Hubbard III (HIII)
approximation, and we here refer to it thus (noting that `resonant
broadening' contributions are additionally included in
Ref. \onlinecite{r13}). HIII is equivalent to UHF, but with a saturated local
moment. Thus, with $|\mu|=1$, the resultant cubic equation for
$G^0(\tw.)$ on the $d^\infty$ BL, obtained from Eqs
(\ref{26},\ref{27}) above, coincides precisely with the HIII
approximation for any $U$; see eg Eq. (34) of
Ref. \onlinecite{r21}. Although Hubbard's original formalism is very
different, its physical content is that of a static approximation to
an alloy analogy description \cite{r22}; a close relationship to UHF
is thus to be expected.

The second connection is to the Falicov-Kimball (FK) model \cite{r14},
a simplified version of the Hubbard model in which electrons of only
one spin type are mobile, and which for $d^\infty$ is exactly soluble
\cite{r23,r24,r25}. For the paramagnetic phase of the FK model the single
particle Green function reduces precisely to that of HIII for any $U$
\cite{r22}, ie to the above-mentioned cubic for $G^0(\tw.)$ on the
$d^\infty$ BL; see eg Eqs (7.3) and (4.4) of Ref. \onlinecite{r25}.

For the P phase, Fig. 1 shows the UHF $D^0(\tw.)$ for $U/t_\ast=4$ on
the $d^\infty$ BL, contrasted to its AF counterpart. Ordering or
otherwise of the preformed local moments clearly has a significant
effect on the spectra. In the AF ordered case, for example, the interior
edges of the Hubbard bands have characteristic square-root
divergences, while for the P phase all band edges vanish with
square-root behaviour. More significantly, while the 2-sublattice
structure of the N\'eel ordered state ensures a band gap
$\Delta=U|\mu|$ for all $U>0$, the single-particle UHF gap vanishes in
the P phase at a critical $U_c\simeq 1.9t_\ast$ given by
$U_c|\mu(U_c)|=\sqrt{2}t_\ast$ ---or correspondingly
$U_c=\sqrt{2}t_\ast$ for HIII/FK--- signalling an insulator--metal
transition. UHF/HIII fails of course in the metallic phase, there
being no well-defined Fermi surface or quasiparticles \cite{r22,r26}. This is
inevitable for any inherently static approximation with a
frequency-independent self-energy $\st._{\alpha\s.}$, 
since the essence of Fermi liquid
behaviour is the inelasticity of electron scattering near the Fermi
level, $\tw.=0$ \cite{r27}.

However, even in the P  insulating phase of interest here, UHF/HIII is 
deficient. As for the AF this is seen most clearly in strong coupling, 
$U\rightarrow\infty$, where although the centres of the Hubbard bands are 
separated by $U$, each has a non-vanishing width. In the strong coupling P
phase, and for the one hole (doublon) sector corresponding to the lower 
(upper) Hubbard band, the Hubbard model maps onto the $t$-$J$ model (Eq.
(\ref{29})) in a {\em random} spin background; and the exact full bandwidth of 
either band is given for the $d^\infty$ BL by \cite{r28,r29}
\begin{mathletters}
\label{210}
\be.
W_\infty=2\sqrt{2}t_\ast~~~~~~; ~U\rightarrow\infty~~.
\ee.
Note that this is also the single-particle bandwidth in the other extreme
of the non-interacting limit, reflecting physically that in strong
coupling the hole/doublon behaves essentially as a free particle
\cite{r30}.

In contrast, the strong coupling bandwidth at UHF/HIII level is 
\be.
W_\infty^0=2t_\ast~~~~~~~~: ~{\rm UHF/HIII~~.}
\ee.
\end{mathletters}
UHF or HIII does not therefore give the exact strong coupling limit for
the Hubbard model, contrary to what has been suggested recently \cite{r31};
but, as is clear from the above discussion, gives instead the strong 
coupling limit of the FK model \cite{r25}. The physical origin of Eq.
(\ref{210}b) is however both simple and revealing. Consider again the
upper Hubbard band in strong coupling, and imagine adding an
$\uparrow$-spin electron to a B-type $\da.$-spin site. Within a static
approximation such as UHF/HIII, only the added $\ua.$-spin can hop; and
it can do so in the first instance to any of $\da.$-spin NN's (B-type
sites) only  ---the effective coordination number for which is $Z_{\rm
eff}=\frac{1}{2}Z$. Since $Z_{\rm eff}$ for the propagating $\ua.$-spin
electron is reduced by a factor of 2 below the full coordination number,
and since the bandwidth of the $d^\infty$ BL is proportional to
$\sqrt{Z_{\rm eff}}$, the strong coupling UHF/HIII width is thus
diminished by $\sqrt{2}$ from the corresponding non-interacting value
which, as in Eq. (\ref{210}a) above, is also the exact strong coupling 
limit; Eq. (\ref{210}b) thus results.

The distinction between Eqs (\ref{210}a) and (\ref{210}b) is however
qualitative, and not solely a matter of degree, reflecting the need to
take seriously ---even in strong coupling--- the {\em correlated}
dynamics of the electrons. Whenever, say, an $\ua.$-spin electron is
added to a site occupied by a $\da.$-spin electron, the added
$\ua.$-spin can indeed propagate in the P phase, scattering elastically
off successive neighbouring $\da.$-spins; and as sketched above this is
well captured at UHF/HIII level. But, having added the $\ua.$-spin to a
$\da.$-spin site, the latter can itself clearly hop off the site ---to a
neighbouring $\ua.$-spin site--- leaving behind it a spin-flip on the
original site. The energy cost for the spin-flip is zero, since we are
considering the P phase where, for $d^\infty$, a given spin is equally
likely to be surrounded by $\ua.$ or $\da.$ spins and has as many
$\ua.$ as $\da.$-spin neighbours (whence there is no
`exchange penalty' for a spin-flip). Thus, whether the added $\ua.$-spin
or the $\da.$-spin already present hops off the site, the initially
created doublon propagates  as a free particle \cite{r30}; Eq. (\ref{210}a)
thus results. To describe correctly the electron dynamics, {\em both}
types of process above ---and therefore the interference between them---
must be included. A static approximation such as UHF/HIII cannot handle
this, since such dynamics reside in the frequency dependence of the full
interaction self-energy $\st._{i\s.}(\tw.)$, as considered in \S 3.

\subsection*{2.2 RPA}

In contrast to single-particle spectra ---probing states one hole or
particle away from $\frac{1}{2}$-filling--- RPA probes fluctuations
about the mean-field state, and thus excitations of the
$\frac{1}{2}$-filled state itself. For the insulating phases, with a
gap to charge excitations, {\em transverse} spin excitations are of
lowest energy. These are reflected in the transverse spin polarization
propagators $\Pi_{ij}^{+-}(t)=i\la. T\{S_i^+(t)S_j^-\}\ra.$ and
$\Pi_{ij}^{-+}(t)$, given within RPA by 
\begin{mathletters}
\label{211}
\be.
\bbox{\Pi}^{+-}(\omega)=\ ^0\bbox{\Pi}^{+-}(\omega)[{\bf 1}-U\
^0\bbox{\Pi}^{+-}(\omega)]^{-1} 
\ee.
where $[\bbox{\Pi}^{+-}(\omega)]_{ij}=\Pi_{ij}^{+-}(\omega)$, $[{\bf
1}]_{ij}=\delta_{ij}$ and $^0\Pi_{ij}^{+-}$ is the pure UHF transverse
spin polarization bubble. Eq. (\ref{211}a) leads directly to a
familiar diagrammatic `bubble sum'. Alternatively, since the
interaction is solely on-site, this may be recast as a `ladder sum' of
repeated particle-hole interactions in the transverse spin channel, as
shown in Fig. 2; bare UHF propagators are denoted by solid lines, and
the on-site interactions (conserving spin at each vertex end) by
wiggly lines. From the basic symmetries (Eq. (\ref{28})), it follows
that $\Pi_{ii}^{-+}(\omega)=\Pi_{ii}^{+-}(-\omega)$ for $i=$ A or B;
and $\Pi_{\hbox{\tiny BB}}^{+-}(\omega)=\Pi^{-+}_{\hbox{\tiny
AA}}(\omega)$.

For finite-$d$, intermediate sites in the ladder sum for
$\Pi_{ii}^{+-}$ (Fig. 2) are in general different from $i$. But since
$^0\Pi_{ij}\sim O(d^{-m})$ for sites $i$ and $j$ $m^{\rm th}$ nearest
neighbours, {\em all} intermediate sites in $\Pi_{ii}$ are equal to
$i$ for $d^\infty$; ie $i=i_1=i_2=\cdots=i$ whence $\Pi_{ii}^{+-}$ (or
$\Pi_{ii}^{-+}$) is purely algebraic, viz
\be.
\Pi_{ii}^{+-}(\omega)=\ ^0\Pi_{ii}^{+-}(\omega)/[1-U\
^0\Pi_{ii}^{+-}(\omega)] ~~~~: d^\infty ~~~.
\ee.
\end{mathletters}

The spectral density of transverse spin excitations is naturally
reflected in Im$\Pi_{ii}^{+-}(\omega)$, as now considered for the AF
and P phases. 

\subsubsection*{A. AF phase}

For the AF, Fig. 3a shows Im$\Pi_{\haa.}^{+-}(\omega)$ at $U/t_\ast=4$ for the
$d^\infty$ Bethe lattice. Two distinct features are apparent: a low
frequency spin-flip pole (discussed below), and a high energy
Stoner-like band. The latter consists simply of weakly renormalized
Hartree-Fock excitations across the gap in the mean-field
single-particle spectrum. Spectral density for the Stoner bands does
not therefore begin until precisely $|\omega|=U|\mu|$ (see Fig. 1),
and their maximum density occurs for $|\omega|\simeq U$. This is as
found also for finite-$d$, see eg Ref \onlinecite{r32}.

The central feature in Fig. 3a is a low-$\omega$ pole at $\omega_p$,
located via Eq. (\ref{211}b) from $U\,^0\Pi_{ii}^{+-}(\w._p)=1$, and
occurring for all $U>0$ (Fig. 3a, inset). This is the sole remnant,
for $d^\infty$, of the spin wave-like component of the transverse spin
spectrum studied recently at RPA level \cite{r32}; and which, for finite
$d$ and any $U>0$, is naturally gapless. Physically, the single
spin-flip pole at $\w._p$ reflects the general suppression of spatial
fluctuations for $d^\infty$: it corresponds simply to the energy cost
of flipping a spin in the AF background. This is particularly clear in
strong coupling, where the Stoner bands are eliminated entirely. Here,
as is well known \cite{r20,r33}, the RPA transverse spin spectrum
reduces (for any $d$) to the linear spin wave spectrum of the nearest
neighbour AF Heisenberg model, with exchange coupling
$J_\infty=4t^2/U=2t_\ast^2/ZU$, onto which the $\frac{1}{2}$-filled
Hubbard model maps rigorously. And for $d^\infty$ it is
straightforward to show that the resultant linear spin wave spectrum 
collapses to an Ising-like spin-flip pole at
$\w._p^\infty=ZJ_\infty/2=t_\ast^2/U$. Further, since linear spin wave
theory for the Heisenberg model is exact for $\di.$ \cite{r34}, it
follows that in strong coupling UHF+RPA gives the exact spin
excitation spectrum of the  $\frac{1}{2}$-filled Hubbard model.

The occurrence of the single $\w._p$-pole is robust to further
renormalization of particle-hole lines in $\Pi_{ii}^{+-}(\w.)$, as
discussed in \S 3.3. We stress further that to capture it requires the
full ladder sum of repeated p-h interactions shown in Fig. 2:
retention solely of the `bare' polarization bubble diagram will
clearly not suffice.

The necessity of including the AF spin-flip scale will be evident when
discussing the $T=0$ single-particle spectra, \S 3.2,3. Here, we
illustrate briefly its importance at finite temperature, as reflected
in the N\'eel temperature $T_N(U)$. Molecular field theory is exact
for the Heisenberg model in $\di.$\cite{r35}; thus, in strong coupling,
$T_N=ZJ_\infty/4=\frac{1}{2}\w._p^\infty$. At finite $U$, we expect
$T_N(U)\simeq \frac{1}{2}\w._p(U)$ to yield a good estimate of the
N\'eel temperature in a $U$-regime where thermal properties are
dominated by the low-lying spin-flip excitations. Jarrell and Pruschke
\cite{r36,r37} have obtained the finite-$T$ phase diagram for the $\di.$
hypercubic lattice via quantum Monte Carlo (QMC). The thermal
paramagnetic phase above $T_N(U)$ is found to be metallic for
$U/t_\ast\lesssim 3$ and insulating for $U/t_\ast\gtrsim 3$ (with a
small `crossover' regime); it is thus in the latter region that we
expect $T_N\simeq \frac{1}{2}\w._p$. This is borne out. Fig. 4
shows the QMC $T_N(U)$ for the $\di.$ hypercubic lattice, together
with the corresponding $\frac{1}{2}\w._p(U)$ and the exact strong
coupling asymptote $T_N=\frac{1}{2}\w._p^\infty$. The QMC N\'eel
temperature is indeed well described by $\frac{1}{2}\w._p(U)$ down to
$U/t_\ast\sim 3$.

\subsubsection*{B. P phase}

For the $T=0$ P phase, Fig. 3b shows Im$\Pi^{+-}_{\haa.}(\w.)$ at
$U/t_\ast=4$ for the $\di.$ Bethe lattice. Compared to its AF
counterpart (Fig. 3a) the key difference is that the spin-flip pole
occurs at $\w.=0$, reflecting the fact that the energy cost for a
spin-flip is zero in the paramagnetic insulator, as argued physically
in \S 2.1b. The formal origin of this at RPA  level is seen readily by
noting that the bare transverse spin polarization bubble (Fig. 2,
diagram (a)) is given by
\be.
^0\Pi_{\haa.}^{+-}(\w.)=i\int_{-\infty}^\infty {d\tw.'\over 2\pi}
\ G^0_{\ha.\da.} (\tw.')G^0_{\ha.\ua.}(\tw.'-\w.) ~~~.
\label{212}
\ee.
From Eq. (\ref{24}) for the UHF Green functions, using
$G^0_{\ha.\da.}=G^0_{\hb.\ua.}$, it follows that 
\be.
G^0_{\ha.\da.}(\tw.)G^0_{\ha.\ua.}(\tw.)=-{1\over
U|\mu|}\left(G^0_{\ha.\ua.}(\tw.)-G^0_{\ha.\da.}(\tw.)\right)~~~.
\label{213}
\ee.
Hence, using the spectral representation of $G^0_{\ha.\s.}(\tw.)$,
\be.
^0\Pi^{+-}_{\haa.}(\w.=0)={1\over U|\mu|}\int_{-\infty}^0
d\tw. \ \left [D^0_{\ha.\ua.}(\tw.)-D^0_{\ha.\da.}(\tw.)\right]~~~.
\label{214}
\ee.
Since the UHF local moment $|\mu|=|\mu_0|$ is given by Eq. (\ref{25}),
$^0\Pi_{\haa.}^{+-}(\omega=0)=1/U$; and thus from Eq. (\ref{211}b) the
RPA $\Pi_{\haa.}^{+-}(\w.)$ has a spin-flip pole at $\w.=0$.

Note again, as for the AF case, that the full ladder sum of
particle-hole interactions in the transverse spin channel is required
to capture the $\w.=0$ spin-flip pole. Further, although we have shown
explicitly its existence within RPA, the occurrence of the
zero-frequency spin-flip scale is naturally a general feature of the
$d^\infty$ paramagnetic insulating phase where, locally, the ground
state is a doubly degenerate local moment (as for the single-impurity
Anderson model embedded in an insulating host) \cite{r5}.

For both phases, the evident virtues of the RPA for excitations of the
$\frac{1}{2}$-filled state contrast sharply with the deficiencies of
the single-particle spectra at UHF level, \S 2.1. This itself hints at
what is necessary to describe the single-particle spectra
successfully: single-particle processes must be coupled dynamically to
the transverse spin excitations, reflected in the frequency dependence
of the self-energy. This is now considered.

\section*{3. Green functions}
\addtocounter{section}{1}
\setcounter{equation}{0}

It is helpful to separate the full interaction self-energies
$\st._{\alpha\s.}(\tw.)$ as
\begin{eqnarray}
\label{31}
\st._{\ha.\s.}(\tw.)&=&-\frac{\s.}{2}U|\mu|+\Sigma_{\ha.\s.}(\tw.)
\nonumber\\
\st._{\hb.\s.}(\tw.)&=&\frac{\s.}{2}U|\mu|+\Sigma_{\hb.\s.}(\tw.) ~~,
\end{eqnarray}
where $\Sigma_{\alpha\s.}(\tw.)$ ($\alpha=$A or B) excludes the
first-order UHF-type contribution, and contains the dynamics on which we
want to focus. From Eqs (\ref{22},\ref{23}) for the Bethe lattice, the exact
site-diagonal Green functions are thus given formally by
\begin{mathletters}
\label{32}
\begin{eqnarray}
G_{\ha.\s.}(\tw.)&=&[\tw.+\frac{\s.}{2}U|\mu|-S_{\ha.\s.}(\tw.)-
\Sigma_{\ha.\s.}(\tw.)]^{-1} \\
G_{\hb.\s.}(\tw.)&=&[\tw.-\frac{\s.}{2}U|\mu|-S_{\hb.\s.}(\tw.)-
\Sigma_{\hb.\s.}(\tw.)]^{-1} ~~~.
\end{eqnarray}
Here, the medium self-energy is given for the AF and P phases by
\be.
S_{\alpha\s.}(\tw.)=\cases{
\frac{1}{2}t_\ast^2G_{\bar{\alpha}\s.}(\tw.) & : AF \cr
\frac{1}{2}t_\ast^2G(\tw.) & : P }
\ee.
\end{mathletters}
where the site index $\bar{\alpha}=$B or A for $\alpha=$A or B
respectively; and
\be.
G(\tw.)=\frac{1}{2}[G_{\ha.\s.}(\tw.)+G_{\hb.\s.}(\tw.)]
\label{33}
\ee.
is the total Green function.

As at UHF level, $\ua./\da.$-spin symmetry and particle-hole symmetry
for the corresponding spectral densities imply
\begin{mathletters}
\label{34}
\begin{eqnarray}
D_{\ha.\s.}(\tw.)&=& D_{\hb.-\s.}(\tw.) \\
&=&D_{\ha.-\s.}(-\tw.)
\end{eqnarray}
\end{mathletters}
respectively. For the associated Green functions
$G_{\alpha\s.}(\tw.)=G^+_{\alpha\s.}(\tw.) +G^-_{\alpha\s.}(\tw.)$, a
Hilbert transform of Eqs. (\ref{34}) gives directly
\begin{mathletters}
\label{35}
\begin{eqnarray}
G_{\ha.\s.}^\pm(\tw.)&=& G_{\hb.-\s.}^\pm(\tw.) \\
&=& -G_{\ha.-\s.}^\mp(-\tw.)~~~.
\end{eqnarray}
Thus, from Eq. (\ref{33}),
\be.
G(\tw.)=-G(-\tw.)~~;
\ee.
\end{mathletters}
while from Eq. (\ref{32},\ref{35})
\begin{mathletters}
\label{36}
\begin{eqnarray}
\Sigma_{\hb.\s.}(\tw.)&=&\Sigma_{\ha.-\s.}(\tw.) \\
&=&-\Sigma_{\ha.\s.}(-\tw.)
\end{eqnarray}
\end{mathletters}
and likewise for the $\st._{\alpha\s.}$'s. Eqs  (\ref{35}a) with 
(\ref{33}) shows also that $G(\tw.)$ is correctly independent of spin.

The symmetries reflected in Eqs (\ref{35},\ref{36}) play an important role
in the following analysis. For the P phase, note also the physical
interpretation of Eq. (\ref{33}) for $G(\tw.)$: viewing the
paramagnet in terms of randomly oriented local moments, where a site
is equally likely to likely to be A-type as B-type, we can consider
Eq. (\ref{33}) as a configurationally averaged Green function. This
is a natural alloy analogy interpretation but, unlike the static
approximation to such inherent in UHF or HIII, it is formally exact
since no approximation to the interaction self-energies has thus far
been made.

\subsection*{3.1 Self-consistent renormalization}

Our aim now is to develop a specific approximation to the self-energy
which in particular (a) becomes exact in strong coupling,  ensuring
thereby a controlled limit; and (b) is constructed in renormalized
form, enabling a self-consistent solution for the single-particle
Green functions.

A relevant diagram contributing to $\Sigma_{i\s.}$ is shown in
Fig. 5, employing the same diagrammatic notation as Fig. 2. Using
deliberately a strong coupling terminology, its physical
interpretation is as follows (with $t>0$ for convenience): at $t=0$ a
($\s.=$)$\ua.$-spin electron, say, is added to site $i$, thus creating
a `doublon'; at $t_1>0$ the ($-\s.$=)$\da.$-spin electron already
present on site $i$ hops from $i$ to $j$, and at $t_2>t_1$ an
$\ua.$-spin hops from $j$ to $k$; the entire path is then
retraced. The diagram thus describes motion of the doublon (or hole
for $t<0$) from $i\rightarrow j\rightarrow k$ via a correlated
sequence of alternating spin hops, creating behind it a string of
flipped spins. All ladder interactions of the resultant on-site
particle-hole pair ---which reflect the on-site spin-flip created by
motion of the doublon/hole--- are shown explicitly for site $i$ in
Fig. 5; from which it is seen that their sum is exactly
$U^2\Pi_{ii}^{-+}$, with $\Pi_{ii}^{-+}(\w.)$ ($=\Pi_{ii}^{+-}(-\w.)$)
the RPA transverse spin propagator discussed in \S 2.2 (cf Fig. 2). 

It is precisely correlated dynamics of the sort exemplified by Fig. 5
that we seek to include and generalize in the frequency-dependent
$\Sigma_{\alpha\s.}(\tw.)$. To this end we first define an undressed
(or self-consistent host) Green function by 
\be. 
\label{37}
\cg._{ii;\s.}(\tw.)=[G_{ii;\s.}^{-1}(\tw.)+\Sigma_{i\s.}(\tw.)]^{-1}~~~.
\ee.
This is shown diagrammatically in Fig. 6(a), as obtained simply from
Eq. (\ref{37}) using the Dyson equation for the full Green function
$G_{ii;\s.}$ expressed in terms of the UHF propagators and self-energy
insertions. As seen from the figure, the implicit sum over
intermediate sites $j,k$ etc. is thus restricted to exclude site $i$
itself (unlike the full $G_{ii;\s.}$ where the site sums are
free). While including all interactions on sites $j\neq i$,
$\cg._{ii;\s.}$ thus excludes all interactions on site $i$ beyond the
simple first-order UHF contribution to $\st._{i\s.}$. The latter is of
course subsumed into the UHF Green functions (as in \S 2.1), which
here constitute the `bare' propagators; and in this important sense
the above definition, Eq. (\ref{37}), of the host propagator differs
from that of eg Refs \onlinecite{r21,r36,r37,r38,r39,r40} 
(which would be recovered if we set the site local moment $|\mu|=0$).

To generalize the processes contained in Fig. 5, we renormalized the
self-energy as shown in Fig. 6(b), replacing the $-\s.$-spin particle
lines connecting the starred vertices $i$ in Fig. 5 by the
self-consistent host Green function $\cg._{ii;-\s.}$; the infinite set
of diagrams thus retained in $\Sigma_{i\s.}$ follows simply by direct
iteration of Fig. 6(b) using Fig. 6(a) for $\cg._{ii;-\s.}$. This
renormalization is adopted for the following reasons. (i) It ensures
that an on-site spin-flip occurs only when the doublon/hole hops off a
site, and that its outward path is self-avoiding. With reference to
Fig. 5 for example, site $j\neq i$ is guaranteed, likewise $k\neq j$;
while terms with $k=i$ vanish for $\di.$, being at least $O(1/d)$
since $G^0_{ij;\s.}\sim O(d^{-m/2})$ for sites $i$ and $j$ $m^{\rm
th}$ nearest neighbours. (ii) In addition, the resultant site
restrictions further prevent the need to include a class of partially
cancelling exchange diagrams, as illustrated simply in Fig. 7 (where a
sum over $j\neq i$ is implicit). Since $j\neq i$ is guaranteed, the
exchange diagram (Fig. 7(b)) is at least $O(1/d)$ and thus vanishes
for $\di.$, while the `direct' diagram (Fig. 7(a)) is $O(1)$. If,
however, $j=i$ was included in the direct diagram, its exchange
counterpart would also be $O(1)$ and would thus need to be retained.

Our basic approximation to $\Sigma_{i\s.}(\tw.)$ is thus Fig. 6(b),
namely
\begin{mathletters}
\label{38}
\be.
\Sigma_{\ha.\ua.}(\tw.)= U^2\int {d\Omega\over  2\pi i}
\Pi_{\haa.}^{-+}(\Omega)\cg._{\ha.\da.}(\tw.-\Omega)~~~;
\ee.
the remaining $\Sigma_{\alpha\s.}$'s follow by symmetry,
Eq. (\ref{36}). From \S 2.2 the RPA $\Pi_{\haa.}^{-+}(\Omega)$
($=\Pi_{\haa.}^{+-}(-\Omega)$) may be separated into the spin-flip pole
contribution, $Q[\Omega+\w._s-i\eta]^{-1}$ (with pole-weight $Q$),
plus the Stoner contribution; whence Eq. (\ref{38}a) may be cast as 
\be.
\Sigma_{\ha.\ua.}(\tw.)=QU^2\cg._{\ha.\da.}^-(\tw.+\w._s)+\Sigma_{\ha.\ua.}^{\rm
Stoner}(\tw.)
\ee.
with spin-flip frequency:
\be.
\w._s=\cases{ \w._p & : AF \cr
0 & : P }
\ee.
\end{mathletters}
By symmetry, $\cg._{\ha.\da.}(\tw.)=\cg._{\hb.\ua.}(\tw.)$ using Eqs
(3.5--7); and from Eqs (\ref{37},2):
\be.
\label{39}
\cg._{\ha.\da.}(\tw.)=\cases{
[\tw.-\frac{1}{2}U|\mu|-\frac{1}{2}t_\ast^2G_{\ha.\ua.}(\tw.)]^{-1} &
: AF \cr
[\tw.-\frac{1}{2}U|\mu|-\frac{1}{2}t_\ast^2G(\tw.)]^{-1}  & : P }
\ee.

From Eq. (\ref{38}a), $\Sigma_{\ha.\ua.}(\tw.)$ is thus a functional
of the Green functions, the basic equations for which (Eqs (\ref{32}))
must therefore be solved self-consistently, as now described.

\subsection*{3.2 Strong coupling}

We consider first the behaviour in strong coupling, as this can be
extracted analytically. Since $|\mu|=1-O(t_\ast^2/U^2)$, Eqs (\ref{32})
reduce in strong coupling to
\begin{mathletters}
\label{310}
\begin{eqnarray}
G_{\ha.\ua.}(\w.)&=&[\w.-S_{\ha.\ua.}(\w.)-\Sigma_{\ha.\ua.}(\w.)]^{-1} \\
G_{\hb.\ua.}(\w.)&=&[\w.-U-S_{\hb.\ua.}(\w.)-\Sigma_{\hb.\ua.}(\w.)]^{-1}~~~.
\end{eqnarray}
\end{mathletters}
The interaction self-energy, Eq. (\ref{38}), likewise simplifies in
strong coupling, since the Stoner contribution vanishes (see \S 2.2)
and the pole-weight $Q\rightarrow 1$. Hence from Eq. (\ref{38}b),
\be.
\label{311}
U^{-2}\Sigma_{\ha.\ua.}(\w.)=\cg._{\ha.\da.}^-(\w.+\w._s)=\cg._{\hb.\ua.}^-(\w.+\w._s)
~~~.
\ee.
This may be reduced further, noting that $\cg._{\hb.\ua.}^-(\w.)$ is
given by
\be.
\label{312}
\cg.^-_{\hb.\ua.}(\omega)=\int_{-\infty}^{U/2} d\w._1\ {{\cal
D}_{\hb.\ua.}(\w._1) \over \w.-\w._1-i\eta}
\ee.
as a one-sided Hilbert transform of the corresponding lower Hubbard
band spectral density, ${\cal D}_{\hb.\ua.}(\w._1)=-\pi^{-1}{\rm
sgn}(\w._1-U/2){\rm Im}\cg._{\hb.\ua.}(\w._1)$; and from Eqs
(\ref{37},\ref{310}b)
\begin{eqnarray}
\label{313}
\cg._{\hb.\ua.}(\w._1)&=&[\w._1'-U-S_{\hb.\ua.}(\w._1)]^{-1} \nonumber\\
&=& [\w._1'-U]^{-1}+[\w._1'-U]^{-2}S_{\hb.\ua.}(\w._1)+O(U^{-3}) \nonumber\\ 
\end{eqnarray}
where $\w._1'=\w._1+i\eta{\rm sgn}(\w._1-U/2)$. 
From Eq. (\ref{312}),
the leading large-$U$ contribution to $\cg._{\hb.\ua.}^-(\w.)$ thus
arises from the second term in Eq. (\ref{313}), yielding
$\cg.^-_{\hb.\ua.}(\w.)=U^{-2}S_{\hb.\ua.}^-(\w.)$; hence from Eqs
(\ref{311},\ref{32}c):
\be.
\label{314}
\Sigma_{\ha.\ua.}(\w.)=S^-_{\hb.\ua.}(\w.+\w._s)=\cases{
\frac{1}{2}t_\ast^2 G_{\ha.\ua.}^-(\w.+\w._p) & : AF \cr
\frac{1}{2}t_\ast^2 G^-(\w.) & : P }
\ee.

We focus now on the lower Hubbard band (LHB) in strong coupling, viz
$\w.\approx 0 \ll U\rightarrow\infty$; the upper Hubbard band follows
trivially by symmetry. Since
$\Sigma_{\hb.\ua.}(\w.)=-\Sigma_{\ha.\ua.}(U-\w.)$, it follows that
for $\w.$ in the LHB $\Sigma_{\hb.\ua.}(\w.)$ is pure real and
$O(1/U)$; it can thus be neglected. The $G_{\alpha\ua.}^+(\w.\approx
0)$ are likewise pure real, with $G_{\hb.\ua.}^+(\w.)\sim O(1/U)$ and
$G_{\ha.\ua.}^+(\w.)\sim O(1/U^3)$ (as may be shown using
Eq. (\ref{310}) together with the analogue of Eq. (\ref{312}) for
$G_{\alpha\s.}^+(\w.)$); together with
$G^+(\w.)=\frac{1}{2}[G_{\ha.\ua.}^++G_{\hb.\ua.}^+]$, they too may be
neglected. And from Eq. (\ref{310}b),
$G_{\hb.\ua.}^-(\w.)\equiv\cg._{\hb.\ua.}^-(\w.)=U^{-2}S_{\hb.\ua.}^-(\w.)$
which can also be neglected asymptotically. Hence in total,
$G_{\ha.\ua.}(\w.)\equiv G_{\ha.\ua.}^-(\w.)$ and
$G(\w.)=\frac{1}{2}[G_{\ha.\ua.}(\w.)+G_{\hb.\ua.}(\w.)]\equiv
\frac{1}{2} G_{\ha.\ua.}(\w.)$. For the LHB in strong coupling,
Eq. (\ref{310}a) thus reduces to
\begin{mathletters}
\label{315}
\begin{eqnarray}
G_{\ha.\ua.}(\w.)&=&[\w.-{\scriptstyle\frac{1}{2}}t_\ast^2
G_{\ha.\ua.}(\w.+\w._p)]^{-1} ~~~~~~\hbox{: AF} \\
G_{\ha.\ua.}(\w.)&=&[\w.-{\scriptstyle\frac{1}{2}}t_\ast^2
G_{\ha.\ua.}(\w.)]^{-1} ~~~~~~~~~~~~~\hbox{: P .} 
\end{eqnarray}
\end{mathletters}

These are the equations for the corresponding $t$-$J_z$ model on the
Bethe lattice, for an AF and random spin background respectively (see
eg \onlinecite{r11,r41}); the $t$-$J_z$ model itself is naturally equivalent
for $\di.$ to the $t$-$J$ model since the spin excitations are purely
Ising-like. We add in passing that a much more detailed asymptotic
analysis, picking up constant terms $O(1/U)$, leads to the `bare'
$\w.$ in the denominators of Eqs (\ref{315}) being replaced by
$\w.+\w._p$ and $\w.+\frac{1}{2}\w._p$ respectively for the AF and P
phases. These shifts, neglected in the brief analysis above, reflect
simply the presence of the trivial charge terms in the $t$-$J$ model
(see Eq. (\ref{29})); they are irrelevant to our subsequent discussion.

Since the $t$-$J$ limit emerges correctly in strong coupling, the
present theory is thus asymptotically exact. Consider for example the
P phase, noting that for the $U=0$ non-interacting limit the Green
function $G_0(\w.)={\rm Re}G_0(\w.)-i\pi{\rm sgn}(\w.)D_0(\w.)$ is
given by
\begin{mathletters}
\label{316}
\be.
G_0(\w.)=[\w.-{\scriptstyle\frac{1}{2}}t_\ast^2G_0(\w.)]^{-1} ~~~~:~~ U=0 ~~,
\ee.
whence the non-interacting spectrum
\be.
D_0(\w.)={1\over \pi t_\ast}[2-(\w./t_\ast)^2]^{\frac{1}{2}} ~~~~:
~|\w.|\leq\sqrt{2}t_\ast 
\ee.
\end{mathletters}
is a semi-ellipse with full width $2\sqrt{2}t_\ast$. From
Eq. (\ref{315}b) this is also precisely the spectral density for
$G_{\ha.\ua.}(\w.)$ in the lower Hubbard band. And since $G(\w.\simeq
0)=\frac{1}{2}G_{\ha.\ua.}(\w.)$ as above, the total lower Hubbard
band spectrum in strong coupling is $D_L(\w.)=\frac{1}{2}D_0(\w.)$,
see also \S 2.1b; (the normalization factor of $\frac{1}{2}$ naturally
reflects the fact that the remaining half of the single-particle
spectrum occurs in the upper Hubbard band centred on $\w.=U$, viz
$D_U(\w.) = \frac{1}{2}D_0(U-\w.)$). Note further that the Feenberg
(`medium') and interaction self-energies contribute equally to the
$\frac{1}{2}t_\ast^2 G_{\ha.\ua.}(\w.)$ denominator in
Eq. (\ref{315}b) for the P phase.
Physically, this reflects the fact discussed in \S 2.1b that, upon
adding a $\s.$-spin electron to a site, it is equally probable for
either the added $\s.$-spin or the $-\s.$-spin electron already
present to hop off the site. At UHF/HIII level, in contrast, only the
former can by construct occur ($\Sigma_{\alpha\s.}=0$): the analogue
of Eq. (\ref{315}b) is then
$G_{\ha.\ua.}^0(\w.)=[\w.-\frac{1}{4}t_\ast^2
G^0_{\ha.\ua.}(\w.)]^{-1}$, producing an incorrect strong coupling
bandwidth of $2t_\ast$ as argued physically in \S 2.1b.

The AF case itself is discussed further in the following section
since, in contrast to the P phase, the {\em approach} to strong
coupling is subtle and physically revealing. Here we simply add that
(i) in contrast to the P phase, the
$\frac{1}{2}t_\ast^2G_{\ha.\ua.}(\w.+\w._p)$ denominator in
Eq. (\ref{315}a) for the AF stems {\em solely} from the interaction
self-energy $\Sigma_{\ha.\ua.}(\w.)$. Thus at UHF/HIII level atomic
limit behaviour arises (incorrectly), viz $G^0_{\ha.\ua.}(\w.)=1/\w.$,
as argued physically in \S 2.1a. (ii) Although obtained explicitly for
the Bethe lattice, Eq. (\ref{315}a) holds equally for the hypercubic
lattice in strong coupling. This is because retraceable paths, which
by construct are the only self-energy paths for a Bethe lattice, are
for the $\di.$ hypercube also the only paths which restore the N\'eel
spin configuration; see also \onlinecite{r29}.

\subsection*{3.3 Results}

At finite $U$ the basic self-consistency Eqs, (\ref{32}) and
(\ref{38}a), are solved numerically. We consider first the AF
phase. 

\subsubsection*{A. Antiferromagnet}

For $U/t_\ast=10$, Fig. 8 shows the resultant lower Hubbard
band, $D_L(\w.)=\pi^{-1}{\rm Im}G(\w.)$; from particle-hole symmetry
the upper band follows by reflection about the Fermi level,
$D_U(\w.)=D_L(U-\w.)$.

For the same $\w._p$-value (Fig. 3a, inset), Fig. 8 shows also the
corresponding $t$-$J_z$ limit spectrum from Eq. (\ref{315}a). As is
well known \cite{r11} the $t$-$J_z$ spectrum is discrete (and to
illustrate relative intensities is thus shown with height proportional
to integrated weight). Physically, this reflects the fact that the
hole is pinned by the string of spin-flips its motion creates, leading
therefore to spatially localized single-particle excitations and hence
a discrete spectrum; mathematically, it is reflected in convergence of
the continued fraction implicit by iteration of Eq. (\ref{315}a).

Although the $U/t_\ast=10$ spectrum evidently bears a close
resemblance to its $t$-$J_z$ counterpart, it is by contrast
continuous. This persists for any finite $U$: with increasing
interaction strength the individual sub-bands in $D_L(\w.)$ centre
ever closely on their  $t$-$J_z$ counterparts, and their integrated
spectral weights tend to those of the  $t$-$J_z$ limit; but they
retain a finite width, reflecting delocalization of the hole. The
peculiarities of $U\rightarrow\infty$ are further evident in the
$t$-$J_z$ model itself, Eq. (\ref{315}a). For any $\w._p>0$ the
$t$-$J_z$ spectrum is discrete, while for $\w._p=0$ (as in
Eq. (\ref{315}b) for the P phase) the spectrum is continuous: the
point $\w._p=0$ thus corresponds to a transition from localized to
extended single-particle excitations, and since $\w._p\rightarrow t_\ast^2/U$ as
$U\rightarrow\infty$ it is clear that $U=\infty$ is a singular point.

While the physical mechanism leading to delocalization of the hole at
any finite $U$ is not of course inherent in the 
 $t$-$J_z$ model Eq. (\ref{315}a) itself, it is readily
inferred. Consider the N\'eel spin configuration and imagine removing,
say, an $\ua.$-spin electron from an A-type site, $i$. The nearest
neighbours (NN) to any $\ua.$-spin site all all $\da.$-spins. Hence to
leading order in $U$ ---the   $t$-$J_z$ limit--- the hole initially
moves via a NN $\da.$-spin electron hopping onto site $i$, creating 
thereon a spin-flip (with an associated exchange energy penalty);
and the subsequent motion of the hole via such a
correlated sequence of alternating NN spin hops, in leaving behind a
string of upturned spins, would by itself render the hole spatially
confined.

At large but finite $U$ there is however a small but non-vanishing
probability amplitude, of order $t_\ast^2/U$, for an $\ua.$-spin
electron on a {\em second} NN site, also A-type, to hop to site $i$
via an intervening $\da.$-spin site: the hole thus moves two lattice
spacings, to the second NN A-type site. Unlike the `` $t$-$J_z$
processes'' above, this does not entail a spin-flip with concomitant
exchange penalty: the hole moves freely.

This mechanism evidently leads to hole delocalization and, in tandem
with the  $t$-$J_z$ processes, produces the strong coupling
spectrum. Its formal origins reside in the passage from
Eq. (\ref{310}a) to Eq. (\ref{315}a) for the AF lower Hubbard band in
strong coupling, where the Feenberg part of the self-energy
$S_{\ha.\ua.}(\w.)=\frac{1}{2}t_\ast^2 G_{\hb.\ua.}(\w.)$ was
neglected. As seen readily from the asymptotics of \S 3.2, the leading
corrections to Im$S_{\ha.\ua.}(\w.\approx 0)$ are
Im$S_{\ha.\ua.}(\w.)=(t_\ast^2/2U)^2{\rm Im}G_{\ha.\ua.}(\w.)$. It is
these that embody the delocalization described above, and lead to
spectral broadening (contributions to Re$S_{\ha.\ua.}(\w.)$ are
$O(1/U)$ and lead simply to residual energy shifts). Further, note
that since the energetic width of the spectral broadening is naturally
the smallest energy scale in strong coupling, the principal effect on
the `bare'  $t$-$J_z$ spectrum is a small resonant broadening of the
individual  $t$-$J_z$ lines. This is seen in Fig. 8, and becomes
clearer still with further increasing $U$.

To our knowledge, the above mechanism is the only one which can lead
to hole delocalization for the $\di.$ AF in strong coupling; and for
the reasons already given in \S 3.2 applies to the hypercubic as well
as the Bethe lattice. In finite-$d$ it is for example well known that
Trugman paths \cite{r42} lead to hole delocalization for the hypercubic
lattice, but such processes are $O(d^{-4})$ and do not therefore
contribute in $\di.$ \cite{r29}.

As $U$ is decreased, the spectra continue to exhibit essentially strong
coupling behaviour down to modest interaction strengths of
$U/t_\ast\sim 2-3$, and can thus be understood quantitatively starting
from the  $t$-$J_z$ limit. This is shown in Ref. \onlinecite{r7} (see eg
Fig. 3(b) therein).

With further decreasing $U$ however, the spectra evolve continuously
to a weak coupling form that shows no trace of remnant  $t$-$J_z$-like
behaviour. The spectral gap closes only in the non-interacting limit
whence, correctly, the system is an AF insulator for all $U>0$. The
full spectrum $D(\tw.)=D_L+D_U$ is shown in Fig. 9 for $U/t_\ast=1$,
together with the corresponding UHF spectrum to which (as one expects)
it is qualitatively closer, although the single-particle gap
$\Delta_g$ is reduced to 0.42 of the UHF gap $\Delta=U|\mu_o|$.

Two further renormalizations have been performed to check the veracity
of the above results. First, note that although the Green functions
have been obtained self-consistently via Eq.s (\ref{32},\ref{38}), the
single-particle propagators occurring in the RPA $\Pi_{\haa.}^{-+}$
that enters the self-energy kernel Eq. (\ref{38}a), are themselves
bare UHF propagators; see Fig. 2. To ensure the theory is robust, we
have thus additionally renormalized the single-particle lines entering
$\Pi_{\haa.}^{-+}$ in terms of both the (self-consistent) full Green
functions $G_{\alpha\s.}$ and the host Green functions
$\cg._{\alpha\s.}$. The results in either case differ only
quantitatively, and at low $U$, from those just described; see also
below.

The second renormalization concerns the local moment $|\mu|$ which, in
the calculations above, has been set to its UHF value $|\mu_0|$. In
weak coupling, van Dongen \cite{r43} has examined perturbatively the
N\'eel temperature and the moment magnitude $|\mu|$ (the order
parameter) for the $\di.$ hypercubic lattice, and has shown that even
for $U\rightarrow 0+$ these are reduced by a factor $q$ of order unity
($q\simeq 0.28$\cite{r43}) below their corresponding UHF values. The
present theory is not of course perturbative (eg the emergence of the
AF spin-flip scale is intrinsically non-perturbative), but it is
certainly closer in spirit to van Dongen to renormalize the moment
beyond UHF level. This is quantitatively important at low $U$, and is
achieved by requiring that $|\mu|$ be determined fully
self-consistently via (cf Eq. (\ref{25}))
\be.
\label{317}
|\mu|=\int_{-\infty}^0 d\tw.\ \left[D_{\ha.\ua.}(\tw.)-D_{\ha.\da.}(\tw.)\right]
\ee.
where $D_{\ha.\s.}$ is the full (as opposed to UHF) spectral
density.

For illustration Fig. 9 shows the Bethe lattice spectrum at
$U/t_\ast=1$, obtained with both $|\mu|$ and $\Pi_{\haa.}^{-+}$
renormalized (the latter in terms of the full Green functions). The
gap $\Delta_g$ is further diminished, the ratio $g=\Delta_g/\Delta$
being $\sim 0.15$; while the local moment $|\mu|$ is likewise reduced
below its UHF counterpart, such that $m=|\mu|/|\mu_0|\sim 0.39$. It is
not unfortunately feasible to obtain numerically accurate estimates of
$g$ and $m$ as $U\rightarrow 0$ (since $|\mu|$ and $\Delta$ rapidly
become exponentially small). But for $U/t_\ast=1$ the UHF moment
itself is accurately represented by its asymptotic $U\rightarrow 0$
limit, $|\mu_0|=8\sqrt{2}{\rm exp}[-\pi t_\ast/\sqrt{2}U-1]$, so the
above result for $m$ may be reasonably close to its limiting value.

\subsubsection*{B. Paramagnet}

To obtain correctly the strong coupling limit for either phase is, as has been
shown, fairly subtle. But in contrast to the AF, the {\em approach} to
strong coupling for the paramagnetic phase is not. Fig. 10 shows the
full spectrum $D(\tw.)=-\pi^{-1}{\rm sgn}(\tw.){\rm Im}G(\tw.)$
($\tw.=\w.-U/2$) for the P phase at $U/t_\ast=8$, 6 and 4, compared to
the strong coupling $t$-$J_z$ limit from Eq. (\ref{315}b). For
$U/t_\ast=8$,  
the strong coupling
limit has in practical terms been reached: the Hubbard bands are
essentially symmetrically centred on $\tw.=\pm U/2$ respectively, with
widths $W\sim W_\infty=2\sqrt{2}t_\ast$ and a band gap of
$\Delta_g\sim\Delta_g^\infty=U-2\sqrt{2}t_\ast$; even for $U/t_\ast=6$
the departure from the asymptotic spectrum is relatively minor. With
further decreasing $U$, however, the individual bands become
increasingly asymmetric; and the gap tends to zero more rapidly than
$\Delta_g^\infty$, signalling the collapse of the insulating
phase. This we now discuss, adding that throughout the insulating
regime the local moments are well developed ($|\mu|\gtrsim 0.95$), as
in Mott's conception of a Mott insulator \cite{r44}.

\section*{4. Destruction of the Mott insulator}
\addtocounter{section}{1}
\setcounter{equation}{0}

Fig. 11 shows the resultant band gap, $\Delta_g(U)$, for the
paramagnetic insulator as a function of $U/t_\ast$. $\Delta_g(U)$ is
found to vanish continuously at a critical $U_c=3.41t_\ast$.
Detailed numerical analysis shows the corresponding exponent to be
unity,
\be.
\label{41}
\Delta_g(U)\sim (U-U_c)^\nu ~~~~~~~:~ \nu=1~~~,
\ee.
and we note that the width of the critical regime is quite narrow: the
behaviour Eq. (\ref{41}) is seen clearly for $(U-U_c)\lesssim 0.05
t_\ast$, corresponding to gaps $\Delta_g(U)\lesssim 0.1t_\ast$. 

The continuous closure of the gap is intimately connected to the
divergence of low-frequency dynamical characteristics of the
system. Consider first the self-energy $\st._{\ha.\ua.}(\tw.)$. At
frequencies $\tw.\in [-\tw._+,\tw._+]$ inside the spectral gap
$(\Delta_g=2\tw._+)$,
$\st._{\ha.\ua.}(\tw.)\equiv\st._{\ha.\ua.}^{\hbox{\tiny  R}}(\tw.)$ is
pure real with a leading low-$\tw.$ expansion
\be.
\label{42}
\st._{\ha.\ua.}^{\hbox{\tiny  R}}(\tw.)-A=B\tw. ~~~~~~~:~ \tw.\rightarrow
0 ~.
\ee.
Here, $A\equiv\st._{\ha.\ua.}(\tw.=0)$
($=-\frac{1}{2}U|\mu|+\Sigma_{\ha.\ua.}(0)$, see Eqs (3.1,8)), and is
finite for all $U>U_c$ (see also below). We wish to find the behaviour
of $B=-|B|$ as $U\rightarrow U_c$.

This is obtained by a scaling analysis. Defining $y=\tw./\tw._+$, it
is found that as the gap closes ($\tw._+\rightarrow 0$),
$\st._{\ha.\ua.}(\tw.)-A$ obeys the scaling form 
\be.
\label{43}
\st._{\ha.\ua.}^{\hbox{\tiny  R}}(\tw.)-A=\tw._+^\alpha\  f(y) ~~~~~~~:~
\alpha={\scriptstyle\frac{1}{2}}
\ee.
with exponent $\alpha=\frac{1}{2}$; i.e. for different values of $U$
close to $U_c$, with correspondingly different gaps
$\Delta_g(U)=2\tw._+(U)$, the $\tw.$-dependent functions 
$[\st._{\ha.\ua.}^{\hbox{\tiny  R}}(\tw.)-A]/\tw._+^{1/2}$ plotted in
terms of $y=\tw./\tw._+$, collapse to a `universal' function
$f(y)$. Four points should be noted about the scaling behaviour. (i)
Good scaling is found in practice for gaps $\Delta_g\lesssim
0.1t_\ast$, consistent with the critical regime found above for
closure of the gap. (ii) The scaling is not confined to frequencies
$y\ll 1$ well inside the spectral gap, but encompasses the region of
non-zero spectral density ($|y|>1$), certainly up to $|y|\sim
2$. (Similar scaling with $\alpha=\frac{1}{2}$ naturally occurs for
Im$\st._{\ha.\ua.}(\tw.)$, as follows from Kramers-Kr\"onig; see also
below). (iii) In numerical terms the scaling analysis is sufficiently
accurate to distinguish readily between an exponent of
$\alpha=\frac{1}{2}$ and, e.g., $\alpha=\frac{1}{3}$. (iv) The scaling
function $f(y)$ is a finite, well behaved function of $y=\tw./\tw._+$,
with $f(y)\sim y$ for $y\rightarrow 0$ as is evident from
Eq. (\ref{42}).

From Eqs (\ref{43}) and (\ref{42}) it follows immediately that
$|B|\sim \tw._+^{-1/2}$; i.e.
\be.
\label{44}
|B|\sim \Delta_g^{-\frac{1}{2}} ~~~~~~~:~ \Delta_g\rightarrow 0
\ee.
or $|B|\sim (U-U_c)^{-1/2}$ from Eq. (\ref{41}) (which we have
confirmed by direct calculation of
$B=(\partial\st._{\ha.\ua.}(\tw.)/\partial\tw.)_0$).

The divergence of $|B|$ controls additionally the low frequency
behaviour of Re$G(\tw.)=X(\tw.)$. From Eq. (\ref{35}c),
$X(\tw.)=-X(-\tw.)$, whence its leading low-$\tw.$ behaviour is
\be.
\label{45}
X(\tw.)=\gamma_1\tw. ~~~~~~~: \tw.\rightarrow 0
\ee.
(with $\gamma_1=-|\gamma_1|$). From Eqs (3.1--3) and (3.6b), $G(\tw.)$
may be written generally  as 
\begin{eqnarray}
\label{46}
G(\tw.)&=&\frac{1}{2}\left\{ 
[\tw.-{\scriptstyle\frac{1}{2}}t_\ast^2G(\tw.)
-\st._{\ha.\ua.}(\tw.)]^{-1} \right. \\
& & \left.  ~~~~~+ [\tw.-
{\scriptstyle\frac{1}{2}}t_\ast^2G(\tw.)+\st._{\ha.\ua.}(-\tw.)]^{-1}\right\}
~~~.
\end{eqnarray}
Using Eqs (\ref{42}) and (\ref{45}) on either side of (\ref{46})
enables $|\gamma_1|$ to be related to $|B|$; the result is 
\be.
\label{47}
|\gamma_1|={1+|B|\over A^2-\frac{1}{2}t_\ast^2}~~~.
\ee.
We find that $A^2>\frac{1}{2}t_\ast^2$ for all $U\geq U_c$, whence
the divergence of $|B|$ as $U\rightarrow U_c$ controls that of
$|\gamma_1|$,
\be.
\label{48}
|\gamma_1|\sim \Delta_g^{-\frac{1}{2}} ~~~~~~~:~ \Delta_g\rightarrow 0 ~.
\ee.

This is further confirmed by a scaling analysis of $X(\tw.)$
itself. In direct analogy to that for $\st._{\ha.\ua.}(\tw.)$ above,
$X(\tw.)$ is found to satisfy the scaling form
\be.
\label{49}
X(\tw.)=\tw._+^\alpha\  x(y) ~~~~~~~: \alpha={\scriptstyle\frac{1}{2}}
\ee.
with $x(y)=-x(-y)$, from which Eq. (\ref{48}) in particular
follows. For $|y|>1$ the corresponding spectrum $D(\tw.)$ likewise
shows the same scaling form as expected, with
$D(\tw.)\sim\tw._+^{1/2}\,[y-1]^{1/2}=
[\tw.-\tw._+]^{1/2}$ for $y\sim 1$ close to the
lower edge of the upper Hubbard band; this, combined with the spectral
representation of $|\gamma_1|$,
\be.
\label{410}
|\gamma_1|=2\int_{\tw._+}^{\infty} d\tw.\  {D(\tw.)\over \tw.^2} ~~~,
\ee.
leads again to Eq. (\ref{48}).

It is instructive to compare the above results with those obtained
from both the simple HIII approximation discussed in \S 2.1B, and with
the resonance broadening contributions \cite{r13} additionally
included, which we refer to as HIII$^\prime$. For HIII, $A=-\frac{1}{2}U$ and
$|B|=0$ ---the approximation is purely static. Eq. (\ref{46}) becomes
a cubic for $G(\tw.)$, leading as is well known to $\Delta_g\sim
(U-U_c)^{3/2}$ \cite{r13}. As is clear from Eq. (\ref{47}) with
$|B|=0$, the transition occurs when $A^2(U_c)=\frac{1}{2}t_\ast^2$,
i.e. $U_c=\sqrt{2}t_\ast$, and $|\gamma_1|\sim (U-U_c)^{-1}\sim
\Delta_g^{-2/3}$. The HIII$^\prime$ approximation can also be shown to be of
the form Eq. (\ref{46}), but with a $\tw.$-dependent
$\st._{\ha.\ua.}(\tw.)$ given by
$\st._{\ha.\ua.}(\tw.)=-\frac{U}{2}+t_\ast^2G(\tw.)$ at low
frequencies (which is sufficient to analyze the critical behaviour);
so that $A=-\frac{U}{2}$ and $|B|=t_\ast^2|\gamma_1|$. Since
$\st._{\ha.\ua.}(\tw.)$ is a simple linear function of $G(\tw.)$,
Eq. (\ref{46}) again becomes a cubic for $G(\tw.)$; and, as for HIII,
the gap exponent $\nu=\frac{3}{2}$ \cite{r13}. Eq. (\ref{47}) with
$|B|=t_\ast^2|\gamma_1|$ yields
$|\gamma_1|=(A^2-\frac{1}{2}t_\ast^2)/(A^2-\frac{3}{2}t_\ast^2)$. The
transition thus occurs when $A^2(U_c)=\frac{3}{2}t_\ast^2$,
i.e. $U_c=\sqrt{6}t_\ast$ as is well known \cite{r13}; and, again,
$|\gamma_1|\sim (U-U_c)^{-1}\sim \Delta_g^{-2/3}$.

Both HIII and HIII$^\prime$ are thus in the same universality class, reflected
more generally in the fact that in either case scaling of the form
Eq. (\ref{49}) can be shown to hold, but with an exponent of
$\alpha=\frac{1}{3}$. Gros \cite{r44a} has recently extended Hubbard's
hierarchical equation of motion decoupling scheme to higher order. The
critical exponents are unchanged from those of HIII/HIII$^\prime$; and the
value of $U_c$ itself is barely changed from its HIII$^\prime$ value of
$U_c/t_\ast\simeq 2.45$. From the above discussion it is apparent that
the present theory belongs to a different universality class from  that
of HIII or its extensions. 

In direct analogy to the AF phase discussed in \S 3.3, we have tested
the robustness of our results by further self-consistently
renormalizing single-particle lines in the polarization propagator
$\Pi_{\haa.}^{-+}(\w.)=\ ^0\Pi_{\haa.}^{-+}/(1-U\ ^0\Pi_{\haa.}^{-+}$)
that enters the self-energy kernel, Eq. (\ref{38}a). To illustrate
what this involves, consider renormalizing $^0\Pi_{\haa.}^{-+}$ (and
hence $\Pi_{\haa.}^{-+}$) in terms of the self-consistent host Green
functions $\cg._{\alpha\s.}$. The resultant $^0\Pi_{\haa.}^{-+}(\w.)$
is then given generally by Eq. (\ref{212}), with the bare (UHF) Green
functions $G^0_{\ha.\s.}$ now replaced by $\cg._{\ha.\s.}$. For
$\w.=0$ in particular, Eq. (\ref{214}) likewise holds, but with the
bare $D^0_{\ha.\s.}(\tw.)$ replaced by the renormalized spectral
densities ${\cal D}_{\ha.\s.}(\tw.)=-\pi^{-1}{\rm sgn}(\tw.){\rm
Im}\cg._{\ha.\s.}(\tw.)$; ie
\be.
\label{49a}
U\ ^0\Pi_{\haa.}^{-+}(\w.=0)=\frac{1}{|\mu|}\int_{-\infty}^0
d\tw.\  [{\cal D}_{\ha.\ua.}(\tw.)-{\cal D}_{\ha.\da.}(\tw.)] ~~~.
\ee.
As discussed in \S 2,3 the key feature of the paramagnetic insulator
is the zero-frequency spin-flip scale. To preserve this, the local
moment $|\mu|$ in Eq. (\ref{49a}) is itself renormalized to ensure that
at each step of the self-consistent iteration scheme $U\,
^0\Pi_{\haa.}^{-+}(\w.=0)=1$ (and we note that throughout the entire
insulating regime, the resultant moment $|\mu|$ is also
self-consistent in the sense of Eq. (\ref{317}) to $< 1\%$
accuracy). The results of this further renormalization are found to
differ negligibly from those we have reported above.

Finally, to demonstrate the importance of the $\w._s=0$ spin-flip
scale, we have eliminated it: both by (a) neglecting its contribution
to $\Sigma_{\ha.\ua.}(\tw.)$ in Eq. (\ref{38}b), retaining only
$\Sigma_{\ha.\ua.}^{\rm Stoner}(\tw.)$; and (b) replacing
$\Pi_{\haa.}^{-+}$ by $^0\Pi_{\haa.}^{-+}$ in the self-energy
kernel Eq. (\ref{38}a). Results obtained from (a) and (b) are very
similar, but differ qualitatively from those reported above. In
particular, although the self-energy remains $\w.$-dependent, the
resultant critical behaviour is found to be that of HIII/HIII$^\prime$
---the gap closes continuously, but with an exponent
$\nu=\frac{3}{2}$. This points clearly to the
necessity of including the $\w._s=0$ spin-flip scale throughout the
entire insulating phase: not only in achieving the correct strong
coupling limit (as in \S 3.2), but also in describing the destruction
of the insulating state.

\section*{5. Discussion}

We now discuss the present work, particularly in relation to the
iterated perturbation theory (IPT) approach
\cite{r21,r38,r39,r40,r45}, use and application of
which has been extensive \cite{r5}. Although our theory of the
Mott-Hubbard insulating phases, with its explicit emphasis on local
moments, is conceptually and technically distinct from IPT, some
general points of marked contrast are evident.

For the antiferromagnetic phase we have emphasized the importance of
the $\w._p$ spin-flip scale, inclusion of which is necessary to obtain
even qualitatively reasonable results throughout essentially the
entire range of interaction strengths, and in particular to recover
exact strong coupling asymptotics. However IPT does not appear to
capture the AF spin-flip scale, presumably because it omits repeated
particle-hole interactions of the sort shown in Fig. 2 (which, as in
\S 2.2A, are required to pick up the spin-flip). This is seen, for
example, from the known inability of IPT to describe correctly the
$U$-dependence of the N\'eel temperature \cite{r5}, particularly in the
`Heisenberg' regime.

For the paramagnetic insulator, the results of \S 4 also disagree
qualitatively with those obtained from IPT; see in particular
\onlinecite{r40} and the review \onlinecite{r5}. Within IPT the
paramagnetic insulating solution is found to break down {\em
discontinuously} (at a critical $U_{c1}=3.67t_\ast$, where the IPT gap
$\Delta_g(U_{c1})\sim 0.3t_\ast$), and $|\gamma_1|$ (Eq. (4.10))
remains {\em finite} at the transition.

The same authors \cite{r12} have recently examined the insulator via
exact diagonalization (ED) on clusters of $n_s=3$, 5 and 7 sites,
extrapolated to $n_s\rightarrow\infty$ assuming $1/n_s$ scaling
behaviour. The resultant data suggest a continuous closure of the gap
at a $U_{c1}/t_\ast=3.04\pm 0.35$ and are consistent with
$\Delta_g(U)\sim (U-U_c)$; see also \onlinecite{r5}. Further, and
independently of the gap analysis, the behaviour of $|\gamma_1|$ has
also been examined by ED \cite{r12}, noting (see Eq. (4.10)) that a
divergence in $|\gamma_1|$ implies a continuous closure of the gap:
$1/|\gamma_1|$ is found to show good scaling behaviour, and to scale
to zero when $n_s$ is extrapolated to $\infty$.

The present theory evidently agrees with the inferences drawn from
ED. These concur with our predictions (\S 4) that the gap closes
continuously and with an exponent $\nu=1$, that $|\gamma_1|$ diverges,
and (less importantly) the value of $U_c$ itself; note moreover that
the ED gap \cite{r12} is in rather good agreement with the present
work over a wide $U$ range. As described in \S 4, inclusion of the
$\omega_s=0$ spin flip scale is central
 in describing the destruction of the Mott
insulator. That IPT appears unreliable close to $U_c$ \cite{r5} thus
suggests an incomplete inclusion of the effects of this spin scale
---which cannot be entirely absent since IPT does give the correct
strong coupling spectrum \cite{r46}--- although in physical terms the
origin of the spin-flip scale within IPT is not transparent.

To conclude, we have developed in this paper a theory for the $T=0$
Mott-Hubbard insulating phases of the $\di.$ Hubbard model,
encompassing both the antiferromagnetic and paramagnetic
insulators. The microscopic perspective it affords hinges on the
importance of low-energy scales for insulating spin-flip
excitations. Their existence is physically natural within the explicit
local moment picture intrinsic to the theory, and inclusion of them is
required not only to obtain the strong coupling limits of the
single-particle spectra ---which are captured exactly--- but more
generally to describe the entire insulating regimes, including  for the
paramagnetic phase in particular the destruction of the Mott insulator.

Let us also note what we have not considered: the metallic state of
the paramagnetic phase. But a glimpse of what is required to describe
the metal within the present framework is evident from Fig. 12. For
$U/t_\ast=3.5$, close to the critical $U_c$ of \S 4, this shows the
spectral density of transverse spin excitations
Im$\Pi_{\haa.}^{+-}(\w.)$ (here obtained, as described in \S 4, with
$^0\Pi_{\haa.}^{+-}$ renormalized in terms of the
$\cg._{\ha.\s.}$). The $\w.=0$ spin-flip pole characteristic of the
paramagnetic insulator is evident, and persists  down to
$U_c$. Clearly, however, the spectral edges of the Stoner-like bands
are themselves approaching $\w.=0$. This they do at $U=U_c$, and for
$U<U_c$ in the metallic phase the insulating spin-flip {\em pole} at
$\w.=0$ is replaced by a {\em resonance} at a small non-zero frequency
$\w.=\w._K$, indicative of the Kondo-like physics known to dominate
the correlated metal \cite{r5,r6}. Extension of the present approach to
describe the metal, encompassing the Kondo spin-scale in such a manner
that the correlated state is correctly a Fermi liquid, will be
described in a subsequent paper.

\acknowledgments

DEL expresses his warm thanks to Ph. Nozi\`eres for the hospitality
of the Institut Laue-Langevin, Grenoble, with particular thanks to
Ph Nozi\`eres, F Gebhard and N Cooper for many stimulating discussions
on the subject matter of this work. MPE acknowledges an EPSRC studentship,
and we are further grateful to the EPSRC (Condensed Matter Physics) for
financial support.

\begin{figure}\caption{
UHF single-particle spectrum, $D^0(\tw.)$, vs $\tw.=\w.-U/2$ (in units
of $t_\ast$) for $\di.$ Bethe lattice. At $U/t_\ast=4$, for AF
phase (full line) and P phase (dashed).
}\end{figure}

\begin{figure}\caption{
Particle-hole ladder sum in transverse spin channel, for RPA
$\Pi_{ii}^{+-}$. Bare (UHF) propagators are denoted by solid lines,
on-site interactions by wiggles. For $d=\infty$, all intermediate sites
$i_1\ldots i_n$ are equal to $i$.
}\end{figure}

\begin{figure}\caption{
Im$\Pi_{\protect\haa.}^{+-}(\w.)$ vs $\w./t_\ast$ at $U/t_\ast=4$ for $\di.$
BL. (a) For AF phase; inset shows $U/t_\ast$ dependence of AF
spin-flip pole $\omega_p$, with dotted line denoting
$U\rightarrow\infty$ asymptote $\w._p^\infty=t_\ast^2/U$. (b) For P
phase, where spin-flip pole $\omega_s=0$ for all $U$ in insulating state.
}\end{figure}

\begin{figure}\caption{
QMC N\'eel temperature vs $U/t_\ast$ (open circles) for $d^\infty$
hypercubic lattice {\protect\cite{r24,r25}}. The simple estimate
$T_N\simeq \frac{1}{2}\w._p(U)$, argued to be valid for
$U/t_\ast{\protect\gtrsim} 3$, is also shown (solid line for $U/t_\ast>3$). The
strong coupling asymptote $T_N^\infty=t_\ast^2/2U$ is indicated by the
dotted line.
}\end{figure}

\begin{figure}\caption{
Diagram contributing to single-particle self-energy $\Sigma_{i\s.}$,
with same notation as Fig. 2; for full discussion, see text.
}\end{figure}

\begin{figure}\caption{
(a) Undressed (or self-consistent host) Green function
$\cg._{ii;\s.}$, expressed in terms of bare (UHF) propagators and
site-diagonal self-energy insertions $\Sigma_{i\s.}$. Note the
restrictions on intermediate site sums. (b) Basic approximation to
$\Sigma_{i\s.}$ used in present work, from which full set of diagrams
retained follows by iteration using Fig. 6(a).
}\end{figure}

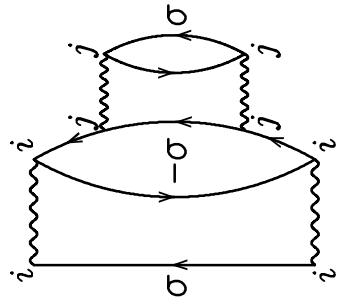
\begin{figure}\caption{
A `direct' diagram (a), and its exchange counterpart (b). With site
$j=i$ excluded from the implicit sum over $j$, the direct diagram is
$O(1)$ while the exchange diagram is $O(1/d)$.
}\end{figure}

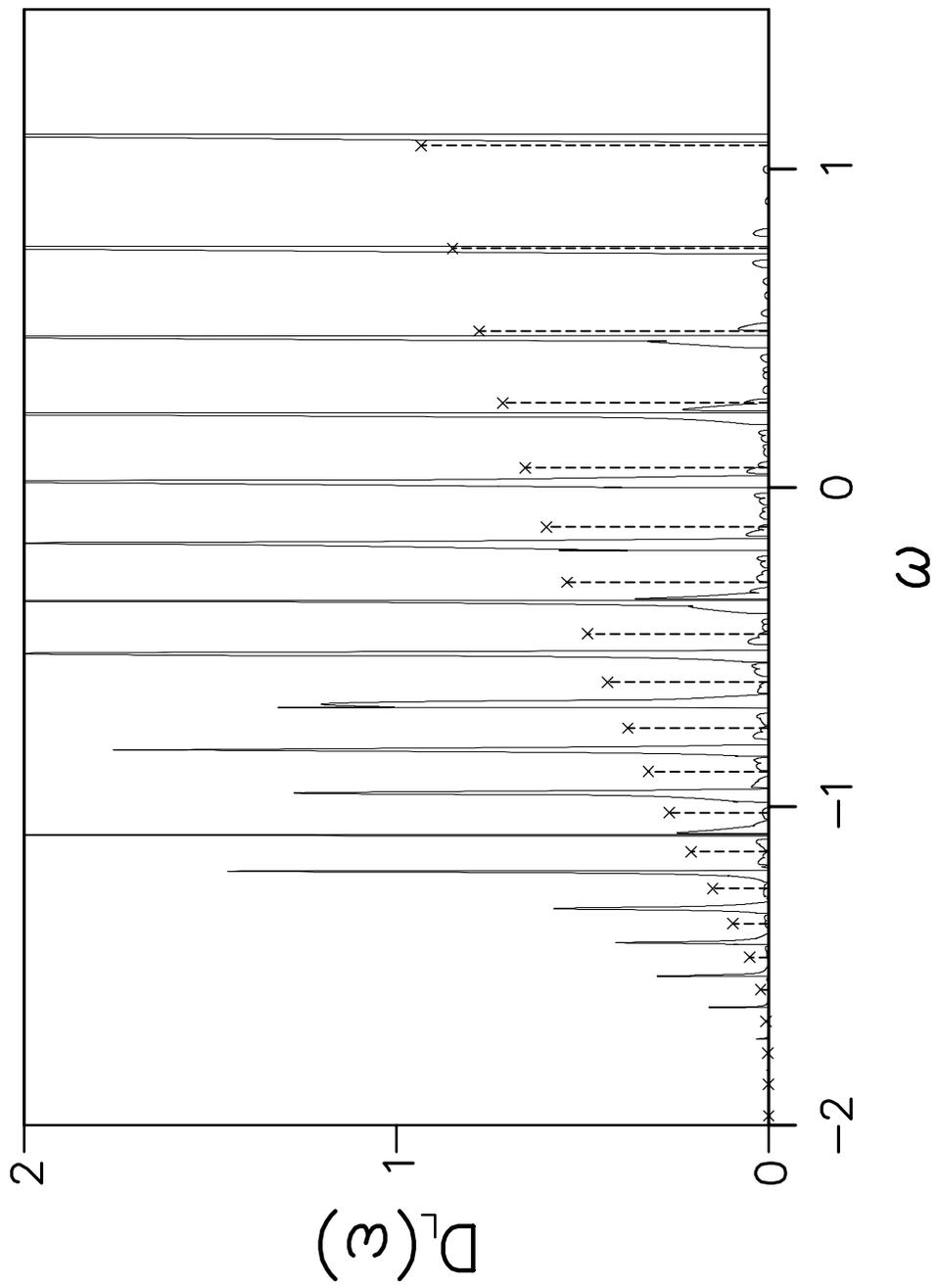
\begin{figure}\caption{
Lower Hubbard band spectrum $D_L(\w.)$ vs $\w.$ (in units of $t_\ast$)
for AF phase (Bethe lattice) at $U/t_\ast=10$; the Fermi level lies at
$U/2=5$. The corresponding $t$-$J_z$ limit spectrum is also shown, as
discussed in text.
}\end{figure}

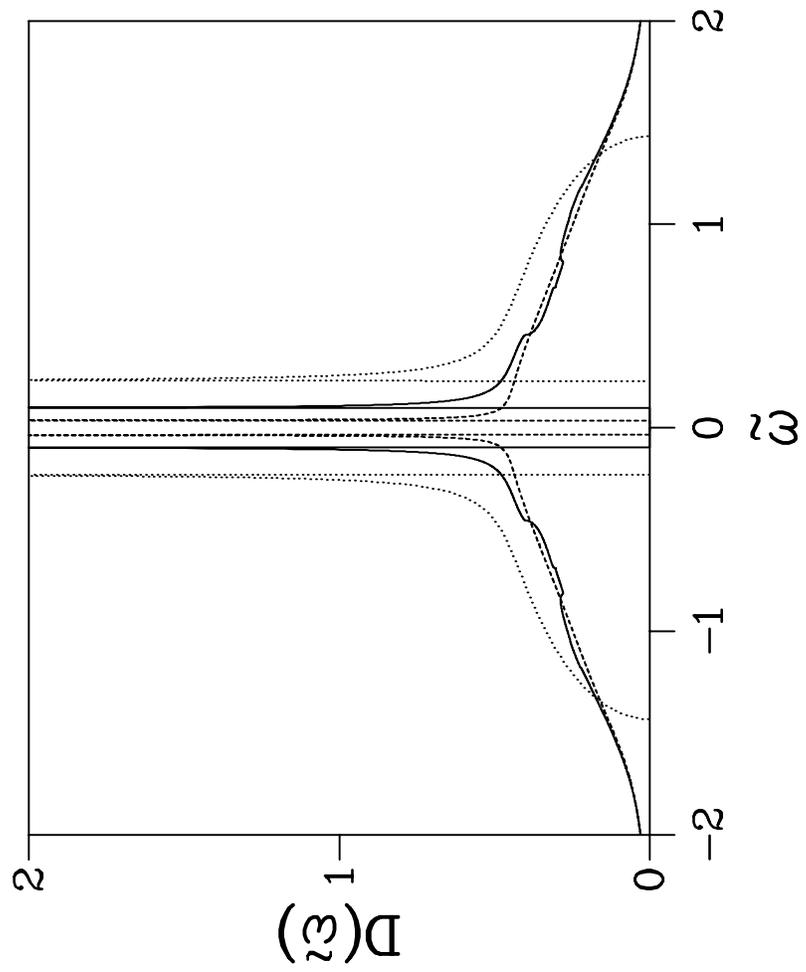
\begin{figure}\caption{
Full spectrum $D(\tw.)$ vs $\tw.=\w.-U/2$ (in units of $t_\ast$) for
AF phase (Bethe lattice) at $U/t_\ast=1$. Dotted line: UHF spectrum;
full line: from present theory; dashed line: with $|\mu|$ and
$\Pi_{\protect\haa.}^{-+}$ further renormalized as described in text.
}\end{figure}

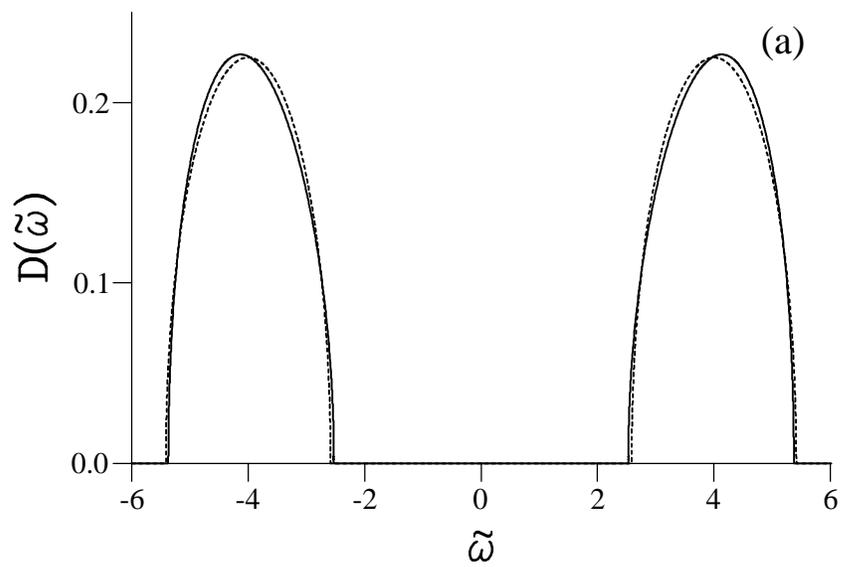
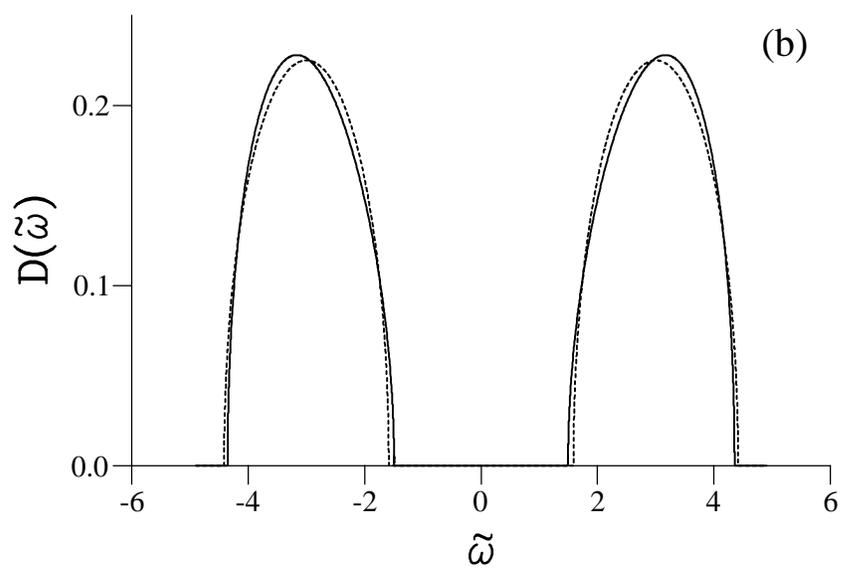
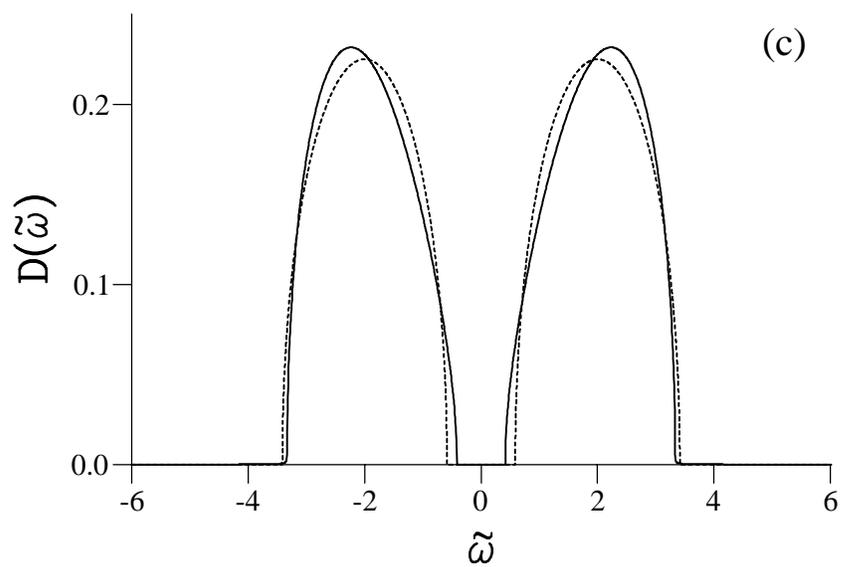
\begin{figure}\caption{
Full spectra $D(\tw.)$ vs $\tw.=\w.-U/2$ (in units of $t_\ast$) for P
phase (Bethe lattice) at $U/t_\ast=8$ (a), 6 (b) and 4
(c). Corresponding strong coupling spectra are shown as dashed lines.
}\end{figure}

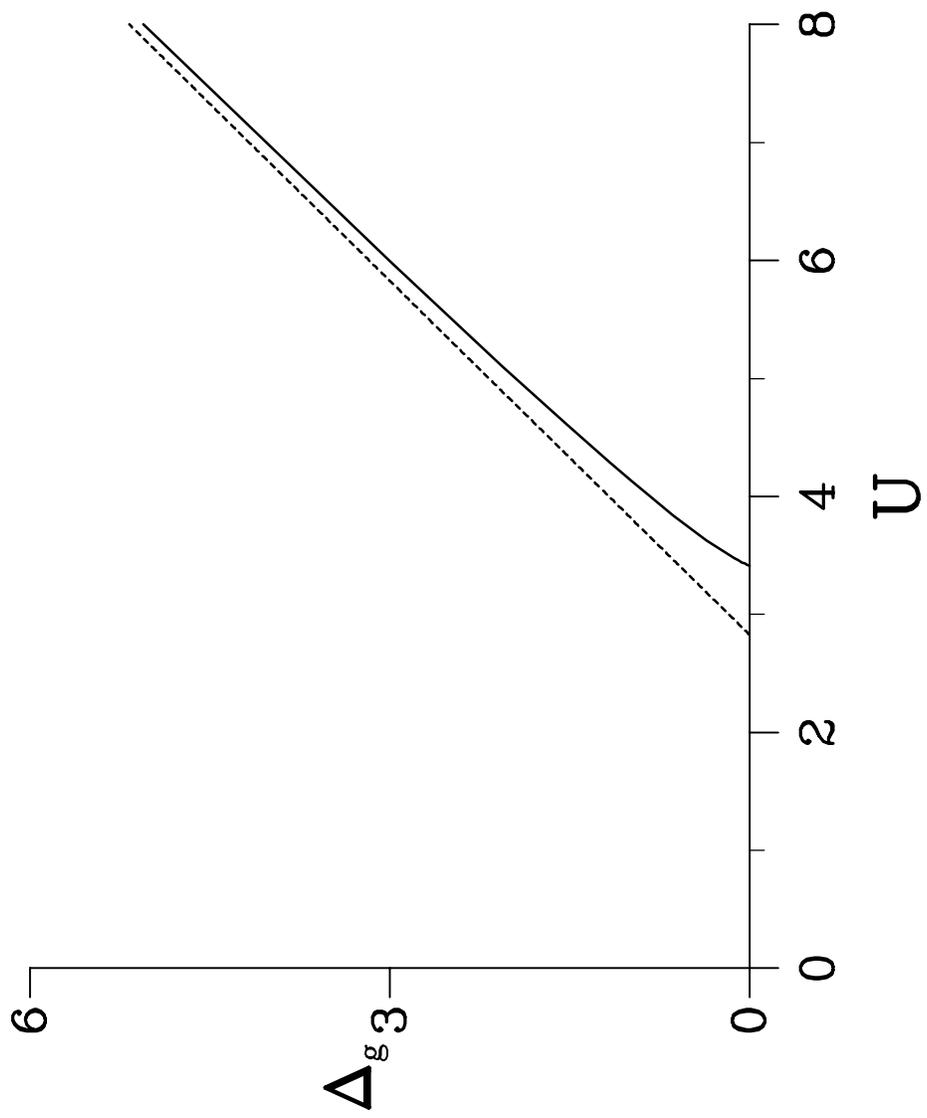
\begin{figure}\caption{
Resultant spectral gap $\Delta_g(U)$ vs $U/t_\ast$ for P insulator
(Bethe lattice). The strong coupling asymptote
$\Delta_g^{\infty}=U-2{\protect\sqrt{2}}t_\ast$ is also shown (dashed line).
}\end{figure}

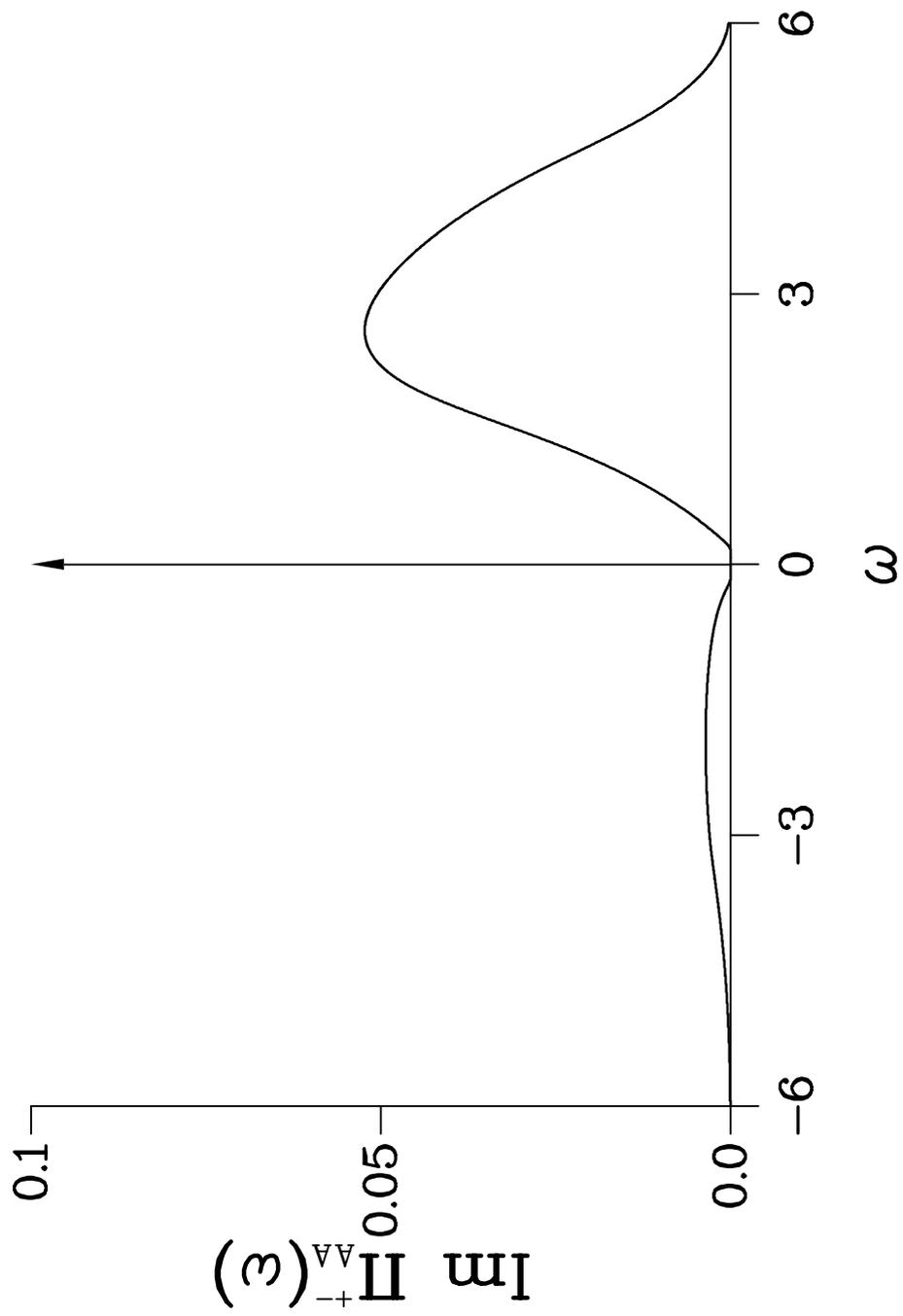
\begin{figure}\caption{
Im$\Pi_{\protect\haa.}^{+-}(\w.)$ vs $\w.$ (in units of $t_\ast$) at
$U/t_\ast=3.5$ close to the boundary of the P insulating state, with
renormalization as described in text.
}\end{figure}

\end{multicols}

\begin{references}
\bibitem{r30} This holds for the Bethe lattice under consideration
because all hole/doublon self-energy paths are retraceable; it does
not hold for the $d^\infty$ hypercube, see Ref. {\protect\cite{r29}}.
\bibitem{r46} This is a subtle point that can be proven
analytically. It bears little relation to the fact {\protect\cite{r5}}
that IPT ---like UHF/HIII--- is exact in the atomic limit.
\end{references}
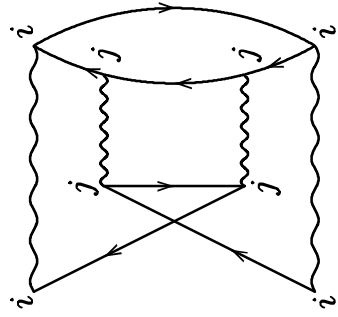
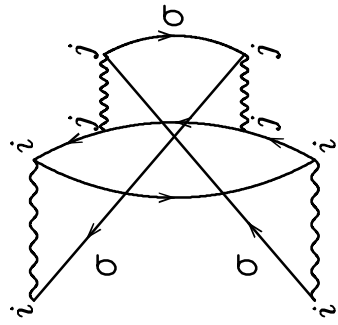
\end{document}